\documentclass[prd,aps,twocolumn,floatfix,preprintnumbers]{revtex4}

\usepackage{epsfig}
\usepackage{amsmath}
\usepackage{latexsym}
\usepackage[psamsfonts]{amssymb}
\usepackage{graphicx}
\usepackage{longtable}
\usepackage{epstopdf}
\usepackage{bm}
\newcommand{\be}{\begin{equation}}
\newcommand{\ee}{\end{equation}}
\newcommand{\bea}{\begin{eqnarray}}
\newcommand{\eea}{\end{eqnarray}}
\newcommand{\bi}{\begin{itemize}}
\newcommand{\ei}{\end{itemize}}

\newcommand{\crg}{c_{g}}
\newcommand{\crf}{c_{f}}

\begin{document}



\title{
Numerical study of tree-level improved lattice gradient flows\\
in pure Yang-Mills theory}


\author{Norihiko Kamata}
\email{kamata@nucl.phys.tohoku.ac.jp}

\author{Shoichi Sasaki}
\email{ssasaki@nucl.phys.tohoku.ac.jp}
\affiliation{Department of Physics, Tohoku University, \\
Sendai 980-8578, Japan}


\date{\today}
\begin{abstract}
We study several types of tree-level improvement in the Yang-Mills gradient flow 
method in order to reduce the lattice discretization errors in line with Fodor {\it et al.} [arXiv:1406.0827].
The tree-level $\mathcal{O}(a^2)$ improvement can be achieved in a simple manner, where
an appropriate weighted average is computed between the plaquette and clover-leaf definitions
of the action density $\langle E(t)\rangle$ measured at every flow time $t$.
We further develop the idea of achieving the tree-level $\mathcal{O}(a^4)$ improvement 
within a usage of actions consisting of the $1\times 1$ plaquette and $1\times 2$ planar loop
for both the flow and gauge actions.
For testing our proposal, we present numerical results for $\langle E(t)\rangle$ obtained 
on gauge configurations generated with the Wilson and Iwasaki gauge actions 
at three lattice spacings ($a\approx 0.1, 0.07,$ and 0.05 fm).
Our results show that tree-level improved flows significantly eliminate the discretization corrections 
on $t^2\langle E(t)\rangle$ in the relatively small-$t$ regime for up to $t\agt a^2$.  
To demonstrate the feasibility of our tree-level improvement proposal, we also study the scaling
behavior of the dimensionless combinations of the $\Lambda_{\overline{\textrm{MS}}}$ parameter
and the new reference scale $t_X$, which is defined through 
$t_X^2\langle E(t_X)\rangle=X$ for the smaller $X$, {\it e.g.}, $X= 0.15$.
It is found that $\sqrt{t_{0.15}}\Lambda_{\overline{\textrm{MS}}}$ shows
a nearly perfect scaling behavior as a function of $a^2$ regardless of the types of gauge action 
and flow, after tree-level improvement is achieved up to $\mathcal{O}(a^4)$. 
Further detailed study of the scaling behavior exposes the presence of the 
remnant $\mathcal{O}(g^{2n} a^2)$ corrections, which are beyond the tree level. 
Although our proposal is not enough to eliminate all $\mathcal{O}(a^2)$ effects, 
we show that the $\mathcal{O}(g^{2n} a^2)$ corrections
can be well under control even by the simplest tree-level $\mathcal{O}(a^2)$ improved flow.
\end{abstract}


\maketitle

 
\section{Introduction}
Recently, the Yang-Mills gradient flow method \cite{{Narayanan:2006rf},{Luscher:2010iy}} has continued to develop remarkably. 
Indeed, this method is extremely useful for setting a reference scale \cite{{Luscher:2010iy},{Borsanyi:2012zs},{Asakawa:2015vta}}, 
computing the nonperturbative running of the coupling constant~\cite{Ramos:2014kla},
defining the energy-momentum tensor (EMT) on the lattice~\cite{{Suzuki:2013gza},{DelDebbio:2013zaa}}, 
calculating thermodynamics quantities \cite{{Asakawa:2013laa},{Kitazawa:2016dsl},{Taniguchi:2016tjc},{Taniguchi:2016lwk}}, and so on~\cite{Luscher:2013vga}. 
These applications are based on measuring the expectation value of the action density $E(t, x)$. 
However, in the calculation of $\langle E(t)\rangle$, there is still room for improvement with respect to the lattice gradient flow, 
where some lattice artifacts are found to be non-negligible~\cite{{Fodor:2014cpa},{Ramos:2015baa}}. 
Therefore, it is important to understand how to practically reduce the effects of the lattice artifacts due to the finite lattice spacing $a$. 

At tree level in the gauge coupling, lattice discretization effects on the expectation value $\langle E(t) \rangle$ 
have been studied in recent years~\cite{{Fodor:2014cpa},{Ramos:2015baa}}. 
According to their results, tree-level discretization errors become large in the small-flow-time ``$t$"
regime as inverse powers of $t/a^2$. 
This tendency is problematic when we construct the lattice EMT operator
using the Yang-Mills gradient flow and then calculate thermodynamics quantities such as the trace anomaly 
and entropy density following Suzuki's proposal~\cite{Suzuki:2013gza}.
The idea of the Suzuki method is based on the fact that flowed observables, which live in $4 + 1$ 
-dimensional space, can be expanded by a series of the expectation values of the ordinary four-dimensional operator 
in powers of the flow time $t$ (the so-called ``small-$t$ expansion'') \cite{Luscher:2013vga}. 
Therefore, it is important to control tree-level lattice discretization errors 
on the action density $E(t, x)$, which is a key ingredient to evaluate the trace anomaly term in the lowest-order 
formula of the new EMT construction~\cite{Suzuki:2013gza}.

In this context, we would like to know what is an optimal combination of choices of the flow, the gauge action,
and the action density in line with the tree-level improvement for the lattice gradient flow~\cite{Fodor:2014cpa}. 
A simple idea of achieving $\mathcal{O}(a^2)$ improvement 
was considered by the FlowQCD Collaboration~\cite{private}. 
The appropriate weighted average of the values $t^2\langle E(t)\rangle$, which are obtained 
by the plaquette and clover lattice versions of $E(t,x)$, can easily cancel their $\mathcal{O}(a^2)$ corrections.
The weight combination was determined at tree level in Refs.~\cite{Fodor:2014cpa} and \cite{Ramos:2015baa}. 
We have developed this idea to achieve tree-level $\mathcal{O}(a^4)$ improvement
using the flow action consisting of both the plaquette and rectangle terms in Ref.~\cite{Kamata:2015ymx}.

In our previous work, it was found that although our tree-level improvement program (which is only valid for $t\agt a^2$) 
significantly eliminated the discretization corrections in almost the entire range of $t$, 
some discretization 
uncertainties still remained in the large-$t$ regime~\cite{Kamata:2015ymx}.
In this paper, we include additional numerical simulations with the renormalization group (RG) 
improved gauge action~\cite{Iwasaki:2011np} and then extend our research scope to address
the feasibility of our improvement program and also fully understand the origin 
of the remnant discretization errors found in our previous work.

This paper is organized as follows. In Sec.~\ref{Sec:2}, 
after a brief introduction of the Yang-Mills gradient flow and its tree-level discretization effects,
we describe our proposal where the tree-level improvement 
is achieved up to and including $\mathcal{O}(a^4)$ in a simple manner 
based on Ref.~\cite{Fodor:2014cpa}.
In Sec.~\ref{Sec:3} we show numerical results obtained from
the pure Yang-Mills lattice simulations using two different gauge actions:
the standard Wilson gauge action and the RG-improved 
Iwasaki gauge action~\cite{Iwasaki:2011np}. Section~\ref{Sec:4} gives the details of
the scaling study of the new reference scale defined in the Yang-Mills gradient
flow method. We then discuss the remnant $\mathcal{O}(g^{2n}a^2)$
corrections, which are beyond the tree level. 
Finally, we summarize our study in Sec.~\ref{Sec:5}.

%
%
\begin{table*}[ht]
\caption{The coefficients $C_{2, 4, 6}$ in the tree-level $\mathcal{O}(a^2)$,
$\mathcal{O}(a^4)$, and $\mathcal{O}(a^6)$ terms of
$t^2\langle E(t)\rangle$ on both the Wilson gauge configurations
and the Iwasaki gauge configurations for various types of the gradient flow~\cite{Fodor:2014cpa}. 
The values of the optimal rectangle coefficient for the $\mathcal{O}(a^4)$-improved flows $c^{\rm WG}_{f1, f2}$ and 
$c^{\rm IG}_{f3, f4}$ are given in the text in Sec.~\ref{Sec:2-C}.
In the table, ``plaq-plus-clover" stands for an appropriate weighted average of 
the plaquette- and clover-type energy densities $\langle E\rangle$
with weight factors defined in Eq.~(\ref{eq:weight}).
\label{tab:Tree-level lattice effects}
}
\begin{ruledtabular} 
\begin{tabular}{clcccc}
types of gauge action &
types of gradient flow & types of $\langle E\rangle$ & $C_2$ & $C_4$ & $C_6$\cr
\hline
Wilson ($c_g=0$)&
unimproved Wilson flow ($c_f=0$)& clover &  $-0.0417$      &$-0.0020 $ &$-0.0002$ \cr
&unimproved Iwasaki flow ($c_f=-0.331$)& clover &  $-0.7037$      &$+0.8490$ &$-1.5093$ \cr
&unimproved Symanzik flow ($c_f=-1/12$)& clover &  $-0.2083$      &$+0.0652$ &$-0.0300$\cr
&$\mathcal{O}(a^2)$-imp Wilson flow ($c_f=0$)& plaq-plus-clover & 0 & $+0.0044$& $+0.0014$ \cr
&$\mathcal{O}(a^2)$-imp Iwasaki flow ($c_f=-0.331$)& plaq-plus-clover& 0 & $-0.0395$& $+0.1362$ \cr
&$\mathcal{O}(a^2)$-imp Symanzik flow ($c_f=-1/12$)& plaq-plus-clover& 0 & $+0.0228$& $-0.0053$ \cr
&$\mathcal{O}(a^4)$-imp Wilson-like flow  ($c_f=c^{\rm WG}_{f1}$)& plaq-plus-clover& 0 &0& $+0.0004$ \cr
&$\mathcal{O}(a^4)$-imp Iwasaki-like flow ($c_f=c^{\rm WG}_{f2}$)& plaq-plus-clover& 0 &0& $+0.0272$ \cr
\hline
Iwasaki ($c_g=-0.331$)&
unimproved Wilson flow ($c_f=0$)& clover &  $-0.2623$      &$+0.0937 $&$-0.0480$ \cr
&unimproved Iwasaki flow ($c_f=-0.331$)& clover &  $-0.9243$      &$+1.2569$ &$-2.3872$ \cr
&unimproved Symanzik flow ($c_f=-1/12$)& clover &  $-0.4290$      &$+0.2395$ &$-0.1765$\cr
&$\mathcal{O}(a^2)$-imp Wilson flow ($c_f=0$)& plaq-plus-clover & 0 & $+0.0408$& $-0.0098$ \cr
&$\mathcal{O}(a^2)$-imp Iwasaki flow ($c_f=-0.331$)& plaq-plus-clover& 0 & $-0.2372$& $+0.6897$ \cr
&$\mathcal{O}(a^2)$-imp Symanzik flow ($c_f=-1/12$)& plaq-plus-clover& 0 & $+0.0003$& $+0.0154$ \cr
&$\mathcal{O}(a^4)$-imp Symanzik-like flow  ($c_f=c^{\rm IG}_{f3}$)& plaq-plus-clover& 0 &0& $+0.0156$ \cr
&$\mathcal{O}(a^4)$-imp positive rectangle flow ($c_f=c^{\rm IG}_{f4}$)& plaq-plus-clover& 0 &0& $-0.1033$ \cr
\end{tabular}
\end{ruledtabular} 
\end{table*}
%

\section{Theoretical Framework}
\label{Sec:2}

\subsection{The Yang-Mills gradient flow and its tree-level discretization effects}
\label{Sec:2-A}

Let us briefly review the Yang-Mills gradient flow and its tree-level discretization corrections. 
The Yang-Mills gradient flow is a kind of diffusion equation where the gauge fields $A _\mu(t, x)$ 
evolve smoothly as a function of fictitious time $t$. It is expressed by the following equation:
%
%
\begin{equation}
\frac{dA_{\mu} (t, x)}{dt} = -\frac{\delta S_{YM}[A]}{\delta A_{\mu}(t, x)},
\end{equation}
where $S_{YM}[A]$ denotes the pure Yang-Mills action defined in terms of the flowed gauge 
fields $A_{\mu}(t, x)$. 
The initial condition of the flow equation at $t=0$, $A_\mu(0, x)$, is 
supposed to correspond to the gauge fields of the four-dimensional pure Yang-Mills theory. 
Through the above flow equation, the gauge fields can be smeared out over the sphere with a
radius roughly equal to $\sqrt{8t}$ in the ordinary four-dimensional space-time.
One of the major benefits of the Yang-Mills gradient flow is that correlation functions of the flowed gauge fields 
$A _\mu(t, x)$ have no ultraviolet (UV) divergence for a positive flow time ($t>0$) 
under standard renormalization~\cite{{Luscher:2010iy},{Luscher:2011bx}}. 

To see this remarkable feature, let us consider a specific quantity, 
like the action density $E(t,x)$ that is defined by 
$E(t,x) = \frac{1}{2}{\rm Tr}\{G_{\mu \nu}(t,x) G_{\mu \nu}(t,x)\}$. 
Here, the field strength of the flowed gauge fields is given by 
$G_{\mu \nu}=\partial_\mu A_{\nu}-\partial_\nu A_{\mu}+[A_\mu, A_\nu]$ ($\mu, \nu= 1, 2, 3, 4$) 
in the continuum expression.
Taking the smaller value of $t$ implies the consideration of high-energy behavior of the theory. 
Therefore, the vacuum expectation of $E(t,x)$ in the small-$t$ regime, where the gauge coupling becomes small, 
can be evaluated in perturbation theory. 
In L\"uscher's original paper~\cite{Luscher:2010iy},  $\langle E \rangle$ 
was given at the next-to-leading order (NLO) in powers of the renormalized coupling in the $\overline{\rm MS}$ scheme, while its next-to-NLO (NNLO) correction has recently been evaluated by
Harlander and Neumann~\cite{Harlander:2016vzb}.

The dimensionless combination $t^2\langle E (t) \rangle$ 
is expressed in terms of the $\overline{\rm MS}$ running coupling $g$ at a scale of $q=1/\sqrt{8t}$
for the pure $SU(3)$ Yang-Mills theory:
%
%
\begin{equation}
\label{eq:flowed_energy}
t^2 \langle E (t) \rangle = \frac{3 g^2(q)}{16 \pi ^2} \Bigl[ 1 
+ \frac{k_1}{4\pi} g^2(q) +\frac{k_2}{(4\pi)^2} g^4(q)+  {\mathcal O}(g^6(q)) \Bigr],
\end{equation}
where the NLO coefficient $k_1$ was obtained analytically as $k_1=1.0978$~\cite{Luscher:2010iy}, 
while the NNLO coefficient $k_{2}$ has been evaluated with the aid of numerical integration
as $k_2=-0.982$~\cite{Harlander:2016vzb}. 
Unlike the ordinary four-dimensional gauge theory, 
Eq.~(\ref{eq:flowed_energy}) has no term proportional to $1/(4-d)$, which is divergent in the limit 
of $d\rightarrow 4$, at this order. 
This UV finiteness has been proved not only for the above particular quantity at this given order, 
but also for any correlation functions composed of the flowed gauge fields at 
all orders of the gauge coupling~\cite{Luscher:2011bx}. 

The lattice version of $t^2 \langle E (t) \rangle$ obtained in numerical simulations
shows a monotonically increasing behavior as a function of the flow time $t$ and 
also good scaling behavior with consistent values of the continuum perturbative calculation 
(\ref{eq:flowed_energy}) that suggests the presence of the proper continuum limit~\cite{Luscher:2010iy}. 
The observed properties of $\langle E (t)\rangle$
offer a new reference scale $t_X$, which is given by the solution of the following equation
%
%
\begin{equation}
\left. t^2 \langle E(t) \rangle \right|_{t=t_X}= X,
\label{eq:new_scale}
\end{equation}
where $X=0.3$ was adopted in Ref.~\cite{Luscher:2010iy}, while an alternative choice of $X=0.4$
has been examined~in Ref.~\cite{Asakawa:2015vta}.

Although the standard Wilson action was used for the lattice gauge action in Ref.~\cite{Luscher:2010iy}, 
in this paper we extend the discussion to an improved lattice gauge action~\cite{Luscher:1984xn} 
in a category of actions consisting of the $1\times 1$ plaquettes
and $1\times 2$ planar loops (``rectangles"), 
which are defined in the $(\mu, \nu)$ plane at a site $x$ 
with the gauge link variables $U(x,\mu)$ as follows:
\begin{widetext}
%
%
\begin{equation}
W_{\mu \nu}^{1\times 1}(x)=\frac{1}{3}{\rm Re}{\rm Tr}
\left[
U(x,\mu)U(x+\hat{\mu},\nu)U^{\dagger}(x+\hat{\nu},\mu)U^{\dagger}(x,\nu)
\right]
\end{equation}
and
%
%
\begin{equation}
W_{\mu \nu}^{1\times 2}(x)=\frac{1}{3}{\rm Re}{\rm Tr}
\left[
U(x,\mu)U(x+\hat{\mu},\nu)
U(x+\hat{\mu}+\hat{\nu},\nu)
U^{\dagger}(x+2\hat{\nu},\mu)
U^{\dagger}(x+\hat{\nu},\nu)
U^{\dagger}(x,\nu)
\right],
\end{equation}
\end{widetext}
where $\hat{\mu}$($\hat{\nu}$) represents the unit vector in the direction indicated by $\mu$($\nu$).

The improved actions we use are given by
%
%
\begin{eqnarray}
S_{\rm lat}(U)&=&-\beta\left\{
(1-8c_{\rm rect})\sum_{x, \mu < \nu}W_{\mu \nu}^{1\times 1}(x)\right.
\nonumber \\
&&\left.\qquad\qquad\qquad+ c_{\rm rect}\sum_{x, \mu, \nu}W_{\mu \nu}^{1\times 2}(x)
\right\},
\end{eqnarray}
which contains two parameters: the bare gauge coupling $g_0$ (being 
$\beta=6/g_0^2$) and the rectangle coefficient $c_{\rm rect}$~\cite{Luscher:1984xn}.
Popular choices for the value of $c_{\rm rect}$ yield the standard Wilson action ($c_{\rm rect}=0$), 
the tree-level Symanzik action ($c_{\rm rect}=-1/12$~\cite{Luscher:1984xn}), and 
the RG-improved Iwasaki action ($c_{\rm rect}=-0.331$~\cite{Iwasaki:2011np}), 
respectively.

The associated flow $V_t(x, \mu)$ of lattice gauge fields is defined by
the following equation with the initial conditions $\left.V_t(x, \mu)\right|_{t=0}=U(x,\mu)$:
%
%
\begin{equation}
a^2\dot V_t(x, \mu)=-g_0^2\left\{
\partial_{x,\mu}S_{\rm lat}(V_t)
\right\}V_{t}(x,\mu),
\end{equation}
where $\partial_{x, \mu}$ stands for the Lie-algebra-valued derivative with respect 
to $V_{t}(x,\mu)$ and the dot notation denotes differentiation with respect to the flow time $t$.

For the above class of lattice gradient flows, tree-level discretization errors of $t^2 \langle E (t) \rangle$ 
were already studied in Refs.~\cite{Fodor:2014cpa} and \cite{Ramos:2015baa}. 
According Ref.~\cite{Fodor:2014cpa}, the lattice version of $t^2 \langle E(t) \rangle$ can be expanded 
in a perturbative series in the bare coupling $g_0$ as
%
%
\begin{equation}
\label{eq:flowed_energy_lattice}
t^2 \langle E (t) \rangle_{\rm{lat}} = \frac{3 g_0^2}{16 \pi ^2} \Bigl[ C(a^2/t) + \mathcal{O}(g_0^2) \Bigr].
\end{equation}
The lattice dependence of the tree-level contribution appears in the first term, which
is classified by powers of $a^2/t$ as $C(a^2/t) = 1 +\sum _{n=1} ^{\infty} C_{2n} \cdot a^{2n} /t^n$.
The second contribution of $\mathcal{O}(g_0^2)$ represents quantum corrections beyond the tree level.
Determinations of the coefficients $C_{2n}$ depend on three building blocks: 1) a choice of the 
lattice gauge action for the configuration generation, 2) a choice of the lattice version of the action 
density, and 3) a choice of the lattice gauge action for the flow action. In Ref.~\cite{Fodor:2014cpa}, 
the $\mathcal{O}(a^{2n})$ correction terms were determined up to $C_8$ for various cases of
the three building blocks.

For clarity, we will hereafter use the term {\it ``X flow''} when we adopt {\it the X gauge action for the flow}. 
For example, we say {\it the Wilson flow} and {\it the Iwasaki flow} when we choose the Wilson and Iwasaki
gauge actions for the flow, respectively.

%
%
\begin{table*}[ht]
\caption{
Simulation parameters of four ensembles generated by the Wilson gauge action (WG).
The values of $r_0/a$~\cite{Sommer:1993ce} and the lattice spacing $a$ are
taken from Ref.~\cite{Guagnelli:1998ud}. $N_{\textrm{conf}}$ is the number of gauge configurations.
\label{tab:Wset-up}
}
\begin{ruledtabular} 
\begin{tabular}{lccccccc}
 $\beta$ (Action) & $L^3 \times T$ & $r_0/a$ & $a$ [fm] & $\sim La$ [fm] & $N_{\textrm{conf}}$ 
 &$t_{0.3}/a^2$ (ours) & $t_{0.3}/a^2$ (L\"uscher) \cr
\hline
5.96 (WG) & $24^3\times 48$&5.005(20) & 0.0999(4)   & 2.40 &  100 & 2.7968(62)& 2.7854(62)\cr
6.17 (WG) & $32^3\times 64$&7.042(30) & 0.0710(3)   & 2.27 &  100 & 5.499(13) & 5.489(14)\cr
6.42 (WG) & $48^3\times 96$&10.04(6) & 0.0498(3)   & 2.39 &  100 & 11.242(23) &11.241(23)\cr
6.42 (WG) & $32^3\times 32$&10.04(6) & 0.0498(3)   & 1.59 &  100 & 11.279(82) & N/A\cr
\end{tabular}
\end{ruledtabular}
\end{table*}
%

\subsection{A simple tree-level $\mathcal{O}(a^2)$ improvement}
\label{Sec:2-B}

Following the tree-level improvement program proposed by Fodor {\it et al.}~\cite{Fodor:2014cpa}, 
we consider several improvements of the lattice gradient flow using
two different choices for the rectangle coefficients: $c_{{\rm rect}, g}$ for the configuration generation, and $c_{{\rm rect}, f}$
for the flow. Hereafter, we simply denote these coefficients as $c_g$ and $c_f$.
First of all, we describe a simple method for tree-level $\mathcal{O}(a^2)$ improvement.
Let us consider the $C_2$ coefficient of the $\mathcal{O}(a^{2})$ correction term
with both the plaquette- and clover-type definitions of the action density $E(t, x)$.
The $C_{2}$ coefficients are given for the plaquette ($C_{2p}$) 
and clover ($C_{2c}$) as follows~\cite{Fodor:2014cpa}:
%
%
\begin{eqnarray}
C_{2p} = 2\crf + \frac{2}{3}\crg + \frac{1}{8}, \quad C_{2c} = 2c_f + \frac{2}{3}c_g - \frac{1}{24}.
\end{eqnarray}
Clearly, $C_{2p}\neq C_{2c}$ with the fixed $c_g$ and $c_f$. Therefore,
in order to eliminate tree-level $\mathcal{O}(a^2)$ effects, one can simply 
take a linear combination of two observables, which gives the corresponding $C_{2}$ coefficient as
$C_{2mix} = \alpha_m  C_{2p} + \beta_m C_{2c}$~\cite{private}.
An appropriate combination of the factors $\alpha_m$ and $\beta_m$ can be determined under the 
condition that $C_{2mix}=0$ with the normalization $\alpha_m + \beta_m = 1$ 
so that the coefficient of the leading term is unity,
%
%
\begin{equation}
\label{eq:weight}
 \alpha_m = 1 - 6\biggl(2c_f + \frac{2}{3}c_g + \frac{1}{8} \biggr),\  \beta_m = 6\biggl(2c_f + \frac{2}{3}c_g + \frac{1}{8} \biggr),
\end{equation}
which can eliminate $C_{2mix}$ for any choice of $c_g$ and $c_f$~\cite{Kamata:2015ymx}.
Therefore, the linear combination
%
%
\begin{equation}
\label{eq:combination}
\alpha_m\langle E_{\rm plaq}(t)\rangle + \beta_m\langle E_{\rm clover}(t)\rangle
\end{equation}
has no tree-level $\mathcal{O}(a^2)$ corrections~\cite{Kamata:2015ymx}.
This idea is quite simple, as can be seen for the case of $c_g=c_f=0$ where a weighted average of two observables, 
$\frac{1}{4}\langle E_{\rm plaq}(t)\rangle + \frac{3}{4}\langle E_{\rm clover}(t)\rangle$, 
would achieve tree-level $\mathcal{O}(a^{2})$ improvement~\cite{difference}.

%
%
 \begin{figure*}[ht]
\includegraphics[width=.32\textwidth,clip]{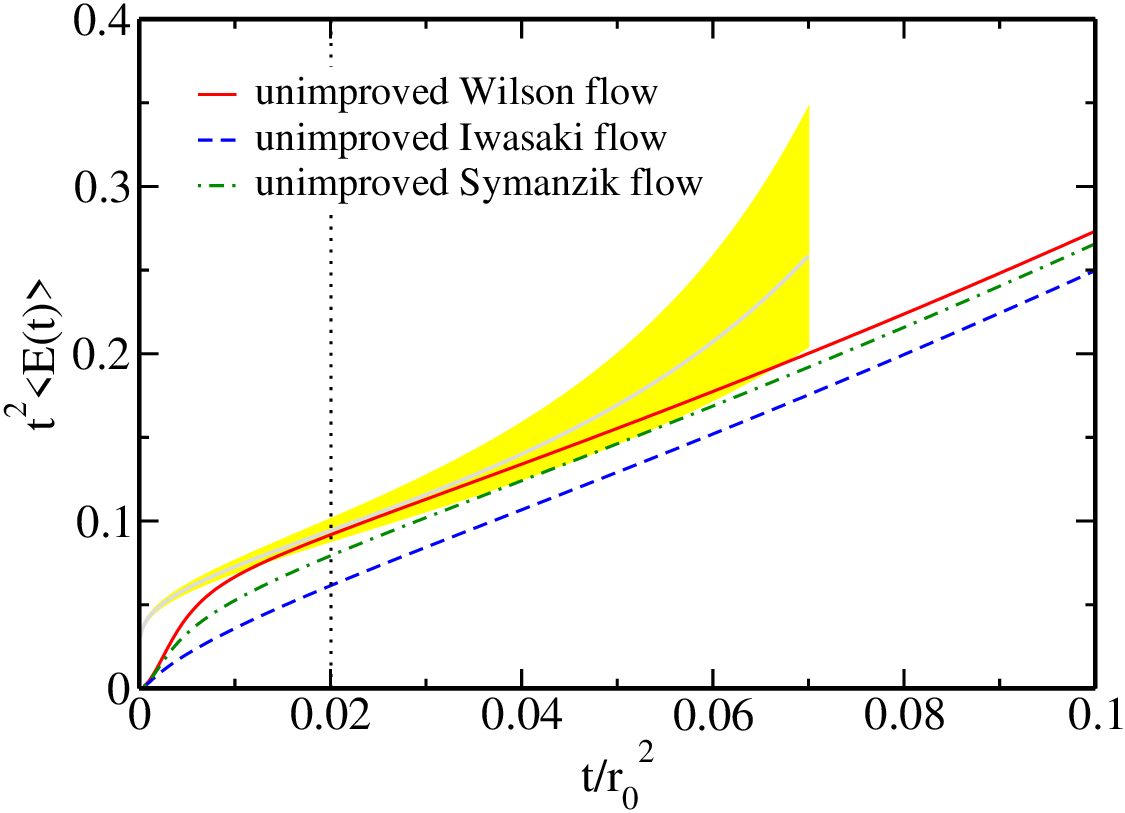}
\includegraphics[width=.32\textwidth,clip]{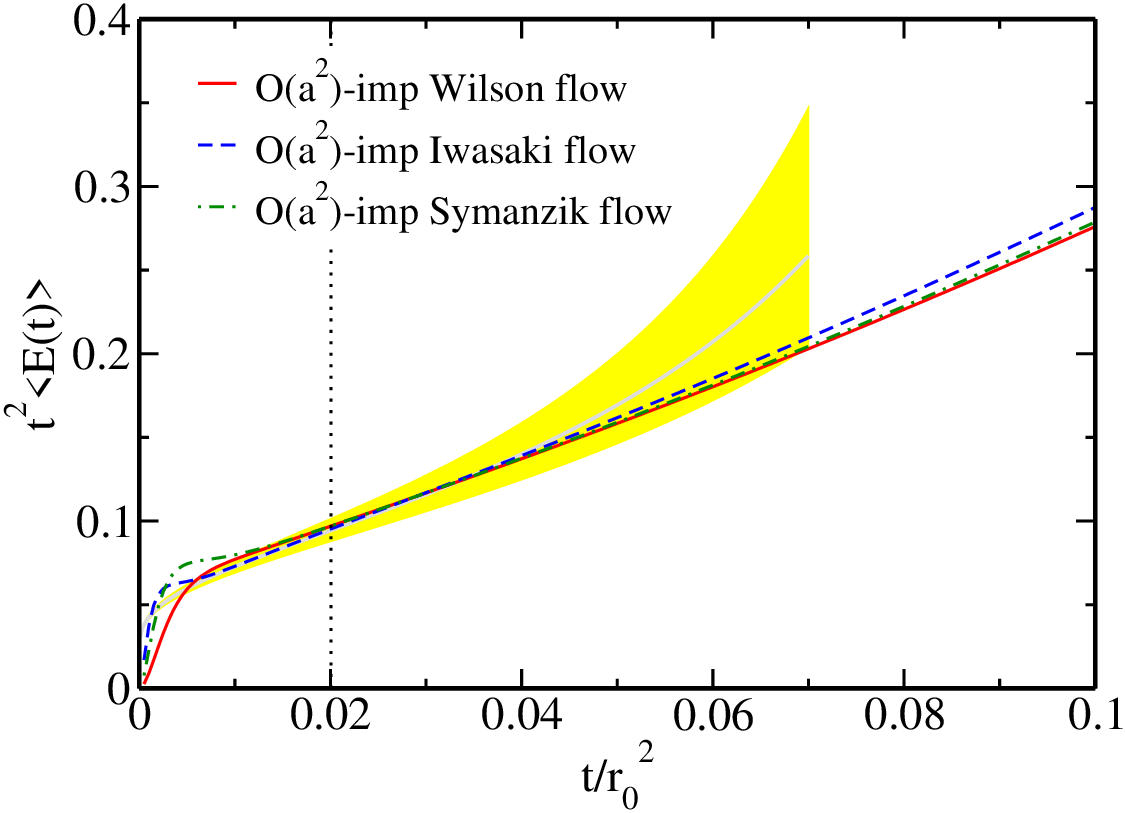}
\includegraphics[width=.32\textwidth,clip]{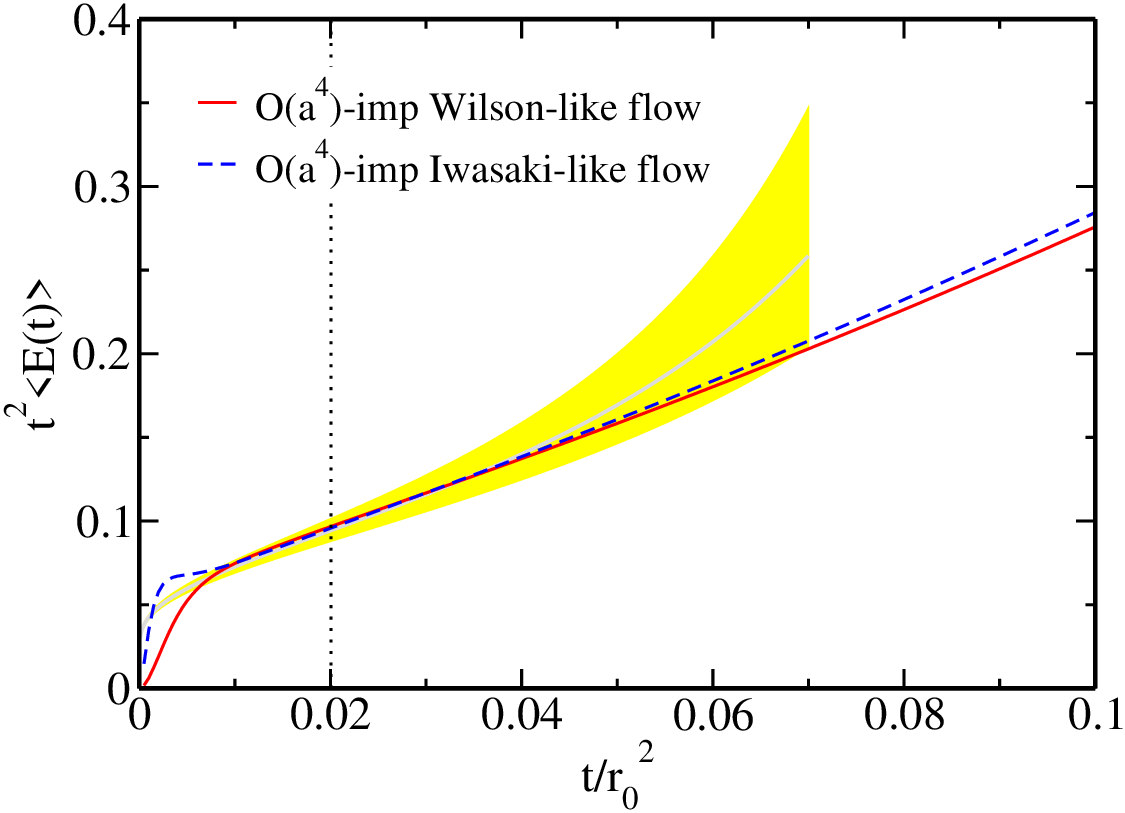}
 \caption{The behavior of $t^2 \langle E (t) \rangle$ calculated on the Wilson gauge configurations at $\beta = 6.17$ as functions of $t/r_0^2$.
The three panels show results for unimproved flows (left), 
their $\mathcal{O}(a^2)$-improved flows (center), and two types of $\mathcal{O}(a^4)$-improved flows (right). 
In each panel, the yellow shaded band corresponds to the continuum perturbative 
calculation~\cite{Luscher:2010iy}. The vertical dotted line in each panel marks the position of $t/r_0^2=a^2/r_0^2$, which corresponds
to the boundary of asymptotic power-series expansions in terms of $a^2/t$. 
}
 \label{fig:beta617}
 \end{figure*}
%

%
%
\begin{figure*}[ht]
\includegraphics[width=.32\textwidth,clip]{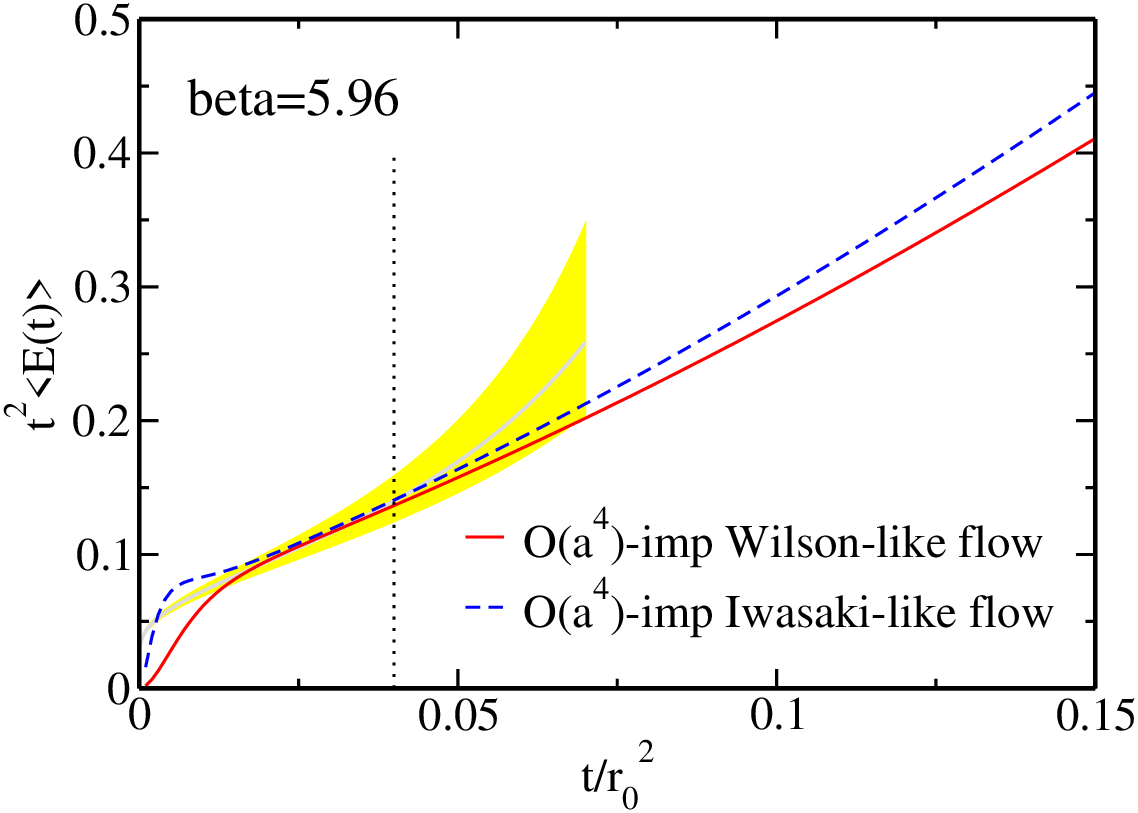}
\includegraphics[width=.32\textwidth,clip]{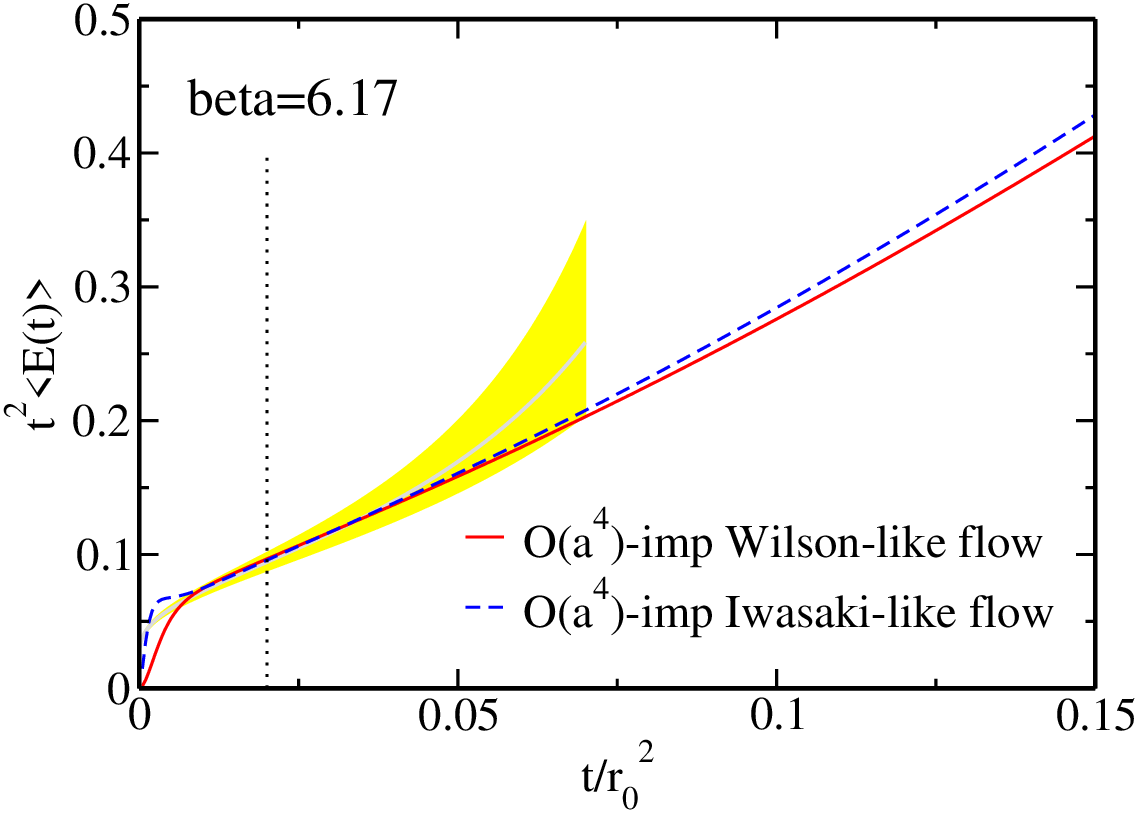}
\includegraphics[width=.32\textwidth,clip]{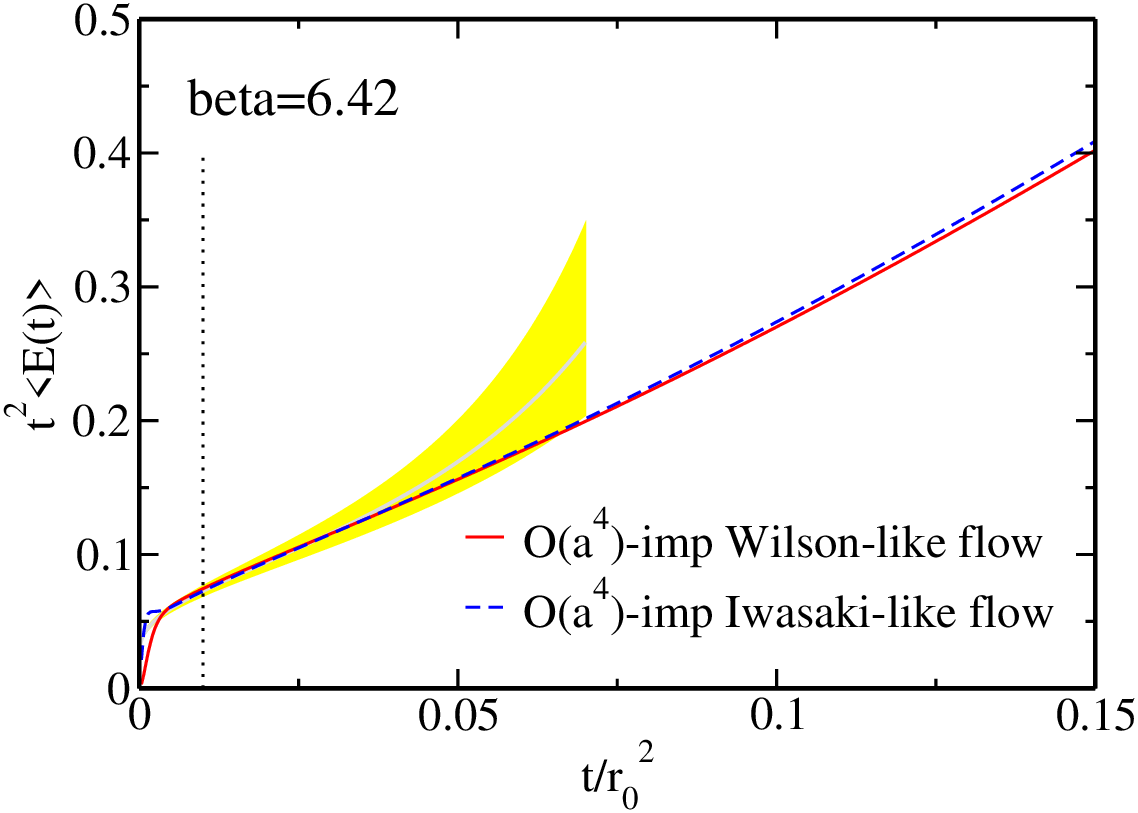}
\caption{The behavior of $t^2 \langle E (t) \rangle$ obtained 
from tree-level $\mathcal{O}(a^4)$-improved flows on the Wilson gauge configurations as functions of $t/r_0^2$.
The three panels show the results calculated at $\beta=5.96$ (left), $\beta=6.17$ (center), and
$\beta=6.42$ (right) (using the same graphical conventions as in Fig.~\ref{fig:beta617}.)}
 \label{fig:o4imp}
 \end{figure*}
%

\subsection{Tree-level $\mathcal{O}(a^4)$-improved gradient flows}
\label{Sec:2-C}

Next, we would like to develop the aforementioned idea to achieve tree-level $\mathcal{O}(a^{4})$ improvement.
Taking a linear combination of $\alpha_m\langle E_{\rm plaq}(t)\rangle + \beta_m\langle E_{\rm clover}(t)\rangle$,
the corresponding $C_4$ coefficient of the $\mathcal{O}(a^{4})$ correction term is given by
%
%
\begin{equation}
\label{eq:order_4th}
 C_{4mix} = \biggl[1 - 6\biggl(x + \frac{2}{3}c_g\biggr) \biggr]C_{4p} + 6\biggl(x + \frac{2}{3}c_g\biggr) C_{4c},
\end{equation}
where $C_{4p}$ and $C_{4c}$ represent the $C_{4}$ coefficients evaluated for the plaquette- 
and clover-type energy densities. Here we introduce $x = 2c_f + 1/8$ for the sake of the following discussion. 
Indeed, coefficients of the higher-order terms $C_4$, $C_6$, and $C_8$ were 
given as polynomial functions of $x$ in Ref.~\cite{Fodor:2014cpa}.
The explicit forms of $C_{4p}$ and $C_{4c}$ are
%
%
\begin{equation}
C_{4p} = \frac{57}{32}x^2 - \frac{25}{128}x + \frac{57}{40}xz + \frac{57}{80}yz + \frac{1}{8}z + \frac{41}{2048}
\end{equation}
and
%
%
\begin{equation}
C_{4c} = \frac{57}{32}x^2 - \frac{25}{128}x + \frac{57}{40}xy + \frac{57}{80}y^2 + \frac{1}{8}y + \frac{53}{2048},
\end{equation}
where $y = c_g -\frac{1}{4}$ and $z = c_g$~\cite{Fodor:2014cpa}. It is found that
the coefficient $C_4$ is at most quadratic in $x$ 
and the coefficient of the highest polynomial term is identical.
Therefore, we are supposed to solve the following quadratic equation in terms of $x$
so as to eliminate $C_{4mix}$:
%
%
\begin{equation}
C_{4mix} = - \frac{57}{160}x^2 - \biggl( \frac{1368c_g +103}{1280} \biggr)x + \frac{48c_g +41}{2048}  = 0,
\end{equation}
which leads to two kinds of optimal coefficients $c_f$ for any gauge action ($c_g$).
Recall that these eliminate $C_{2mix}$ and $C_{4mix}$ simultaneously with a given rectangle coefficient $c_g$.

In this paper, we consider the Wilson gauge action ($c_g = 0$) and the Iwasaki gauge action ($c_g = -0.331$) for numerical simulations. For these setups, the optimal-flow coefficients $c_f$ are given as
%
%
\begin{equation}
\label{eq:opt_coeff_WG}
  c^{\textrm{WG}}_{f1} = 0.012323, \quad c^{\textrm{WG}}_{f2} =  -0.250261
\end{equation}
for the Wilson gauge configurations~\cite{Kamata:2015ymx} and
%
%
\begin{equation}
\label{eq:opt_coeff_IG}
  c^{\textrm{IG}}_{f3} = -0.083756, \quad c^{\textrm{IG}}_{f4} =  0.342317
\end{equation}
for the Iwasaki gauge configurations.
The superscripts of ``WG" and ``IG" found in Eqs.~(\ref{eq:opt_coeff_WG}) and (\ref{eq:opt_coeff_IG}) 
stand for the Wilson and Iwasaki gauge configurations, respectively.
The second and third solutions ($c^{\textrm{WG}}_{f2}$, $c^{\textrm{IG}}_{f3}$) are close 
to the rectangle coefficient of the Iwasaki gauge 
action ($c_f = -0.331$) and the tree-level Symanzik gauge action ($c_f = -1/12$), while 
the first one ($c^{\textrm{WG}}_{f1}$) is very close to zero, 
which corresponds to the Wilson gauge action ($c_f = 0$).  
Therefore, we call the second and third flows an ``Iwasaki-like flow'' 
and ``Symanzik-like flow," while the first one is called a ``Wilson-like flow.'' 
The remaining one  ($c^{\textrm{IG}}_{f4}$) is called a ``positive rectangle flow" for convenience. 
Their $\mathcal{O}(a^2)$, $\mathcal{O}(a^4)$, and $\mathcal{O}(a^6)$ correction terms $C_{2,4,6}$ 
are summarized in Table~\ref{tab:Tree-level lattice effects}.

In addition to the idea of achieving the tree-level improvement
by means of the linear combination of $\langle E_{\rm plaq}(t)\rangle$
and $\langle E_{\rm clover}(t)\rangle$, the original paper proposed 
the following tree-level improved action density:
\begin{equation}
\label{eq:fodor_imp}
\langle E(t)\rangle_{\rm imp} = \frac{\langle E(t)\rangle_{\rm lat}}{1 + \sum^{m_{\rm max}}_{m=1}C_{2m}(a^2/t)^m},
\end{equation}
where the coefficients $C_{2m}$ have been determined up to $C_8$ 
($m_{\rm max}=4$)~\cite{Fodor:2014cpa}. 
We will make a comparison of our method with the original proposal in the Appendix.

\section{Numerical results}
\label{Sec:3}

In this paper we perform the pure Yang-Mills lattice simulation 
using two different gauge actions: the standard Wilson gauge action ($c_g=0$) 
and the RG-improved Iwasaki gauge action ($c_g=-0.331$).
\subsection{Results from the Wilson gauge configurations}
\label{Sec:3-A}
First of all, we focus only on the results obtained from the Wilson gauge configurations
and will discuss the Iwasaki gauge action results in the next subsection.
The gauge ensembles in each simulation with the Wilson gauge action
are separated by 200 sweeps after 2000 sweeps
for thermalization. Each sweep consists of one heat bath~\cite{Cabibbo:1982zn} 
combined with four over-relaxation~\cite{Creutz:1987xi}
steps. As summarized in Table \ref{tab:Wset-up}, we generate three ensembles 
by the Wilson gauge action ($c_g=0$) with fixed physical volume ($La \approx 2.4$ fm) 
(which corresponds to the same lattice setups as in the original work of the Wilson flow 
done by L\"uscher~\cite{Luscher:2010iy}), and additionally generate the smaller 
volume ensemble ($La\approx 1.6$ fm) at $\beta = 6.42$ so as to study the finite-volume effect.
We have checked our code by determining a reference scale of $t_{0.3}/a^2$ from 
the clover-type energy density, which can be directly compared with 
the results of Ref.~\cite{Luscher:2010iy}, as tabulated in Table \ref{tab:Wset-up}.

In the following discussion, 
we use five different types of flow action for the gradient flow---Wilson, Iwasaki, Symanzik, and two 
$\mathcal{O}(a^4)$-improved flows---and then evaluate $t^2 \langle E (t) \rangle$ 
by means of both the plaquette- and clover-type definitions.
To eliminate $\mathcal{O}(a^2)$ or $\mathcal{O}(a^4)$ corrections from the observable of 
$\langle E (t) \rangle$, we take the appropriate linear combinations of $\langle E_{\rm plaq}(t) \rangle$
and $\langle E_{\rm clover}(t) \rangle$ according to Eqs.~(\ref{eq:weight})
and (\ref{eq:combination}). 
In the case of $\mathcal{O}(a^4)$ improvement, 
the optimal coefficients defined by the formula (\ref{eq:opt_coeff_WG}) are used.
Under these variations, we classify eight different types of the gradient flow result 
on the Wilson gauge configurations. Their $\mathcal{O}(a^2)$, $\mathcal{O}(a^4)$, and $\mathcal{O}(a^6)$ 
correction terms $C_{2,4,6}$ are summarized in Table~\ref{tab:Tree-level lattice effects}.

In Fig.~\ref{fig:beta617}, we first show how our proposal of tree-level improvements
works well regarding the $t$ dependence of $t^2 \langle E (t) \rangle$ calculated at $\beta = 6.17$.
The three panels show results for unimproved flows (left), 
$\mathcal{O}(a^2)$-improved flows (center),
and $\mathcal{O}(a^4)$-improved flows (right). [Hereinafter, the two types of tree-level improved flows are 
called ``$\mathcal{O}(a^2)$-imp flow" and ``$\mathcal{O}(a^4)$-imp flow," respectively.]
The red solid (blue dashed) curve in each panel is obtained from the Wilson-type (Iwasaki-type) flows, 
while the green dot-dashed curves in the left and center panels is from the Symanzik flow.
The yellow shaded band in each panel 
represents the continuum perturbative calculation
using the NLO formula of Eq.(\ref{eq:flowed_energy}) with the 
four-loop $\overline{\textrm{MS}}$ running coupling~\cite{{vanRitbergen:1997va},{Czakon:2004bu}}, 
which is the same prescription adopted in Ref.~\cite{Luscher:2010iy}. 

For the unimproved case (left panel), the Wilson flow result is closest to the continuum perturbative calculation.
It is found that the larger absolute value of the rectangle coefficient $c_f$ in the flow action further 
pushes the result away from the continuum perturbative calculation.
This could be caused by the size of tree-level discretization errors of $t^2 \langle E (t) \rangle$,
which has been evaluated in Ref.~\cite{Fodor:2014cpa} 
(as partly summarized in Table~\ref{tab:Tree-level lattice effects}).

Both tree-level $\mathcal{O}(a^2)$ and $\mathcal{O}(a^4)$
improvements indeed significantly improve results obtained from both the Iwasaki-type and Symanzik flows.  
Even for the Wilson-type flows, the improvements become visible in the relatively small-$t$ 
regime up to $t/r_0^2=a^2/r_0^2\approx 0.02$, which 
corresponds to the boundary of asymptotic power-series expansions in terms of $a^2/t$ at $\beta=6.17$. 
Furthermore, it is observed that in the range of $0.02< t/r_0^2 < 0.05$, 
curves obtained from each flow almost coincide. This tendency is likely to be
strong in results for the tree-level $\mathcal{O}(a^4)$-imp flows (right panel) 
especially toward smaller values of $t$. 
This indicates that the tree-level discretization errors, which may dominate in the small-$t$ regime, 
are well controlled by our proposal. However, in the large-$t$ regime ($t/r_0^2>0.05$), 
the difference between results from the Wilson-type flow ($c_f\approx 0$) and 
the Iwasaki-type flow ($c_f\approx -0.3$) becomes evident and also increases 
for a larger value of $t$. 
It is worth mentioning that at tree level, the higher-order corrections become negligible 
in the large-$t$ regime due to powers of $a^2/t$. 

What is the origin of the observed difference appearing in the larger $t$ region? 
There are two major sources. One is the finite-volume effect, which could be different
between the results obtained from different flow actions.  
As mentioned before, the Yang-Mills gradient flow is a kind of diffusion equation, and 
then the radius of diffusion becomes large as the flow time increases. 
Therefore, the flowed gauge fields in the larger $t$ region are 
more sensitive to the boundary of the lattice. However, as we will show later, this is not the case.  
Another possibility is that the difference stems from some 
remaining discretization errors beyond the tree-level discretization effects,
since non-negligible $\mathcal{O}(g^{2n} a^2)$ corrections may appear 
in the larger $t$ region where the renormalized coupling $g^2$ becomes large.

To clarify these points, we focus on the results from two types of 
tree-level $\mathcal{O}(a^4)$-imp flow.
Figure~\ref{fig:o4imp} displays how the observed difference between the Wilson-like flow and the 
Iwasaki-like flow in the large-$t$ regime can change when the lattice spacing decreases.
In Fig.~\ref{fig:o4imp}, from the left panel to the right panel, the corresponding values 
of the lattice spacing in our simulations at a given $\beta$ are going from a coarser to a finer lattice spacing. 

\begin{figure}[ht]
\includegraphics[width=.40\textwidth,clip]{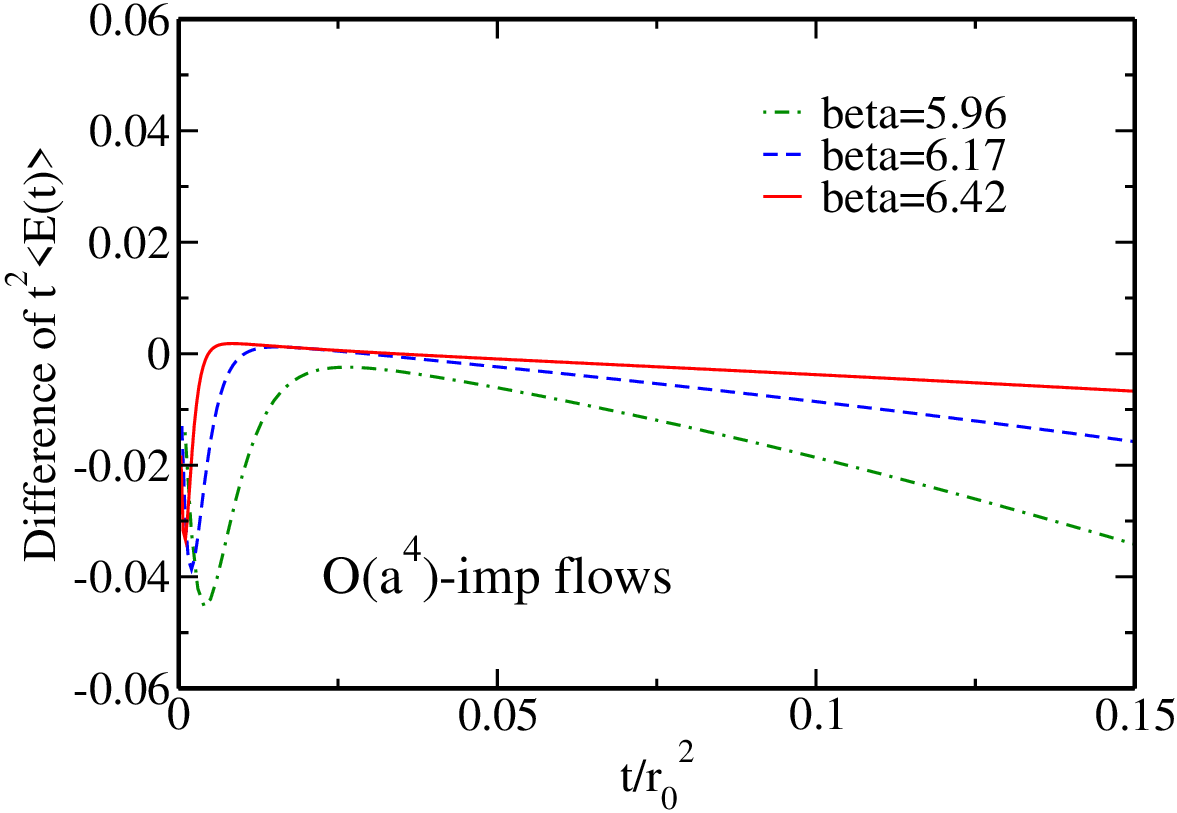}
\caption{Differences in the values of $t^2 \langle E (t) \rangle$ between 
$\mathcal{O}(a^4)$-imp Wilson-like and $\mathcal{O}(a^4)$-imp 
Iwasaki-like flows at different lattice spacings as functions of $t/r_0^2$.
Red solid, blue dashed, and green dot-dashed curves represent results at $\beta=5.96$, 6.17, and 6.42. 
}
\label{fig:WG_t2Edif}
\end{figure}

In Fig.~\ref{fig:WG_t2Edif}, we also plot the differences in the values of $t^2 \langle E (t) \rangle$ between 
the $\mathcal{O}(a^4)$-imp Wilson-like flow and $\mathcal{O}(a^4)$-imp Iwasaki-like flow
as a function of $t/r_0^2$ at each $\beta$. Green dot-dashed, blue dashed, and red solid curves 
denote results at $\beta=5.96$, 6.17, and 6.42. 
This figure clearly shows that the difference, which grows in the larger $t$ region, 
becomes diminished as the lattice spacing decreases. 
Therefore, we confirm that the difference stems from some remaining discretization errors.

\begin{figure}[t]
\includegraphics[width=.40\textwidth,clip]{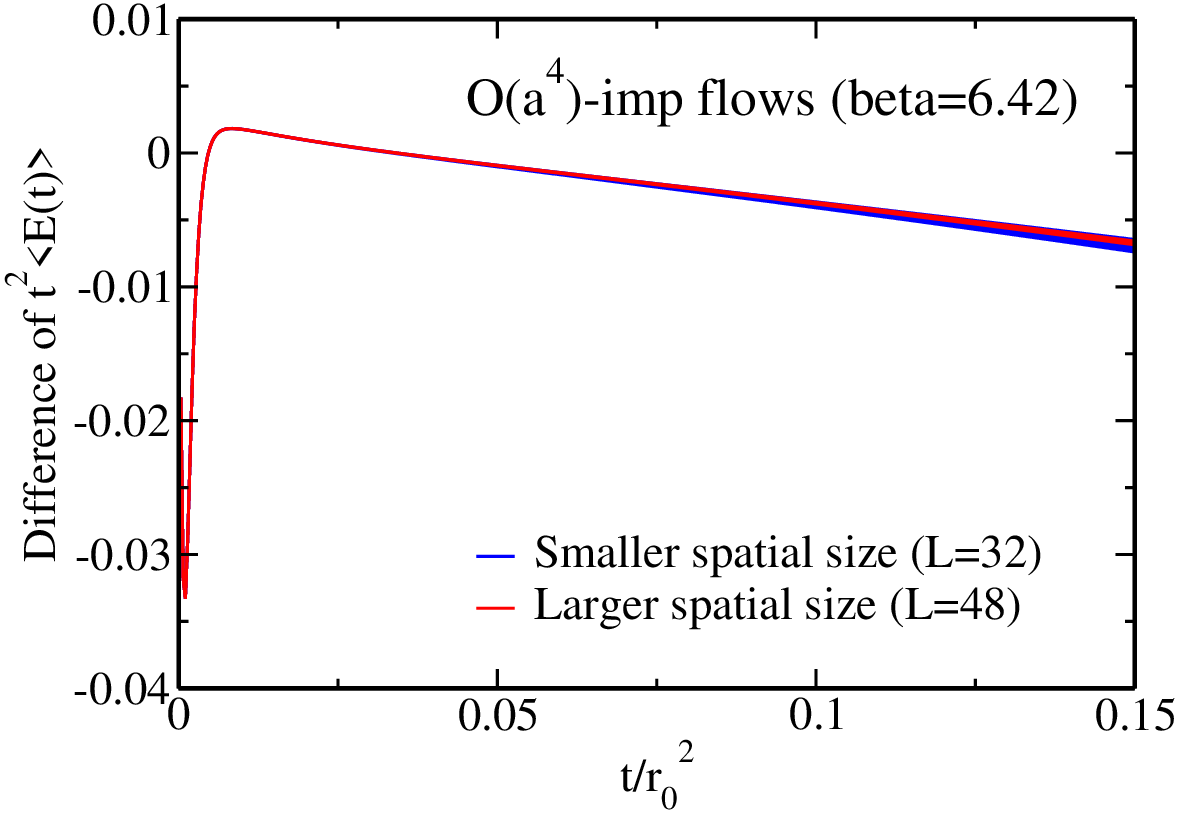}
\caption{Finite-volume dependence in differences in $t^2 \langle E (t) \rangle$ 
between the $\mathcal{O}(a^4)$-imp Wilson-like and Iwasaki-like flows 
as functions of $t/r_0^2$. 
Red and blue shaded bands correspond to the results obtained 
in the larger ($48^3 \times 96$) and smaller ($32^3 \times 32$) volumes, respectively.
}

\label{fig:WG_t2Edif_FV}
\end{figure}

In order to determine the size of the finite-volume effect, we also calculate the differences
in $t^2\langle E(t) \rangle$ between two $\mathcal{O}(a^4)$-imp flow results
on the smaller lattice volumes ($32^3 \times 32$) at $\beta = 6.42$.
Then, we directly compare the results obtained on two different lattice volumes 
($48^3 \times 96$ and $32^3 \times 32$) as shown in Fig.~\ref{fig:WG_t2Edif_FV}.
We confirm that there is no visible finite-volume effect at least in the range of $t/r_0^2 \alt 0.15$. 

From these observations, we conclude that 
the difference between two $\mathcal{O}(a^4)$-imp flows appearing in the larger $t$ region
is caused by non-negligible $\mathcal{O}(g^{2n} a^2)$ corrections beyond the tree-level discretization effects.
We then remark that the reference scale $t_{0.3}$, which is 
determined at around $t/r_0^2\approx 0.1$ (as originally proposed in Ref.~\cite{Luscher:2010iy}),
may suffer from rather large $\mathcal{O}(g^2 a^2)$ errors (of the order of 1\% 
when the lattice spacing is coarse, as large as $a\approx 0.1$ fm).

%
%
%

\subsection{Results from the Iwasaki gauge configurations}
\label{Sec:3-B}

%
%

To evaluate the effectiveness of our proposal, we next study various types
of tree-level improved gradient flows on the gauge configurations generated
by an improved lattice gauge action including the rectangle term.
We choose the Iwasaki gauge action ($c_g=-0.331$) and then generate four gauge ensembles
with similar lattice parameters (spacings $a$ and volumes $La$) to the lattice setups for 
the Wilson gauge action as summarized in Table \ref{tab:Iset-up}. The Iwasaki gauge 
ensembles in each simulation are also separated by 200 sweeps after 2000 sweeps 
for thermalization, as in the cases of the Wilson gauge configurations.
The smaller volume ensemble ($32^3\times 32$) at the finer lattice spacing ($\beta=3.10$) 
is reserved for the finite-volume study. 

We use five different types of flow action for the gradient flow---Wilson, Iwasaki, Symanzik,
and two $\mathcal{O}(a^4)$-imp flows (the same as in Sec.~\ref{Sec:3-B})---for the evaluation
of values of $t^2 \langle E (t) \rangle$. Their $\mathcal{O}(a^2)$, $\mathcal{O}(a^4)$, 
and $\mathcal{O}(a^6)$ correction terms $C_{2,4,6}$ are summarized in 
Table~\ref{tab:Tree-level lattice effects}.

%
%
\begin{figure*}[ht]
\includegraphics[width=.32\textwidth,clip]{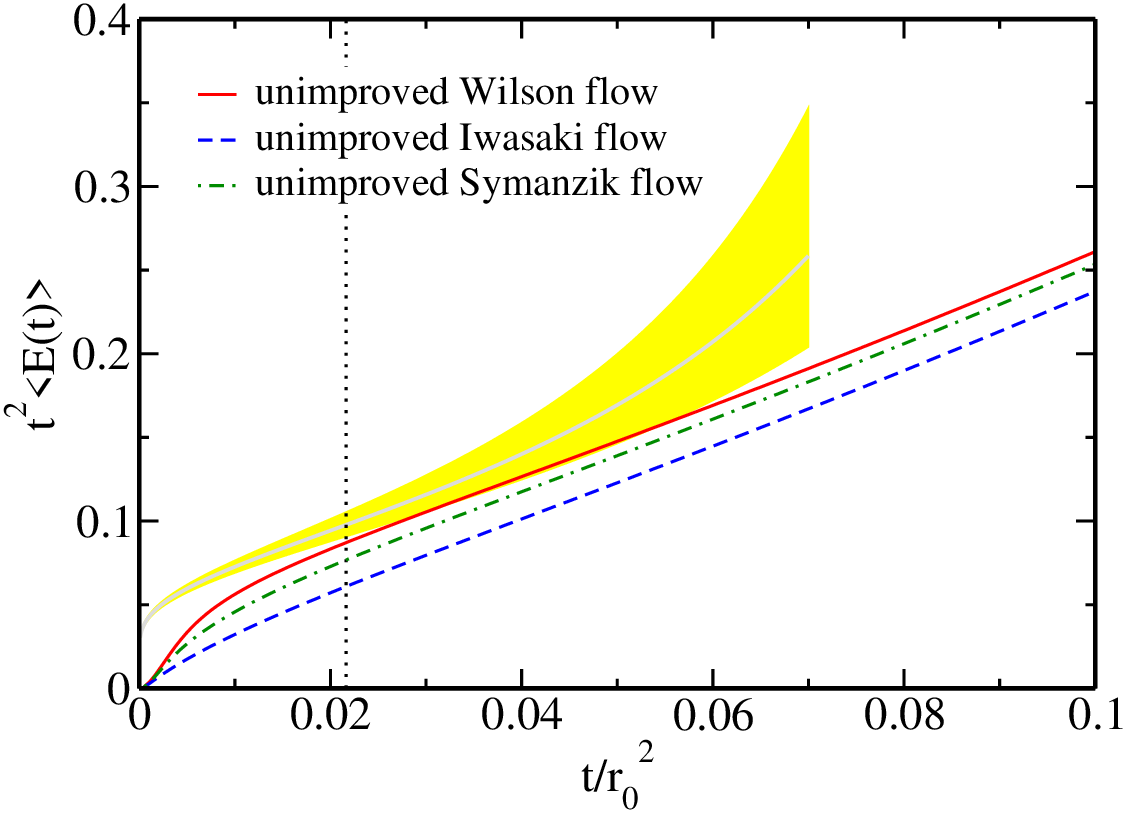}
\includegraphics[width=.32\textwidth,clip]{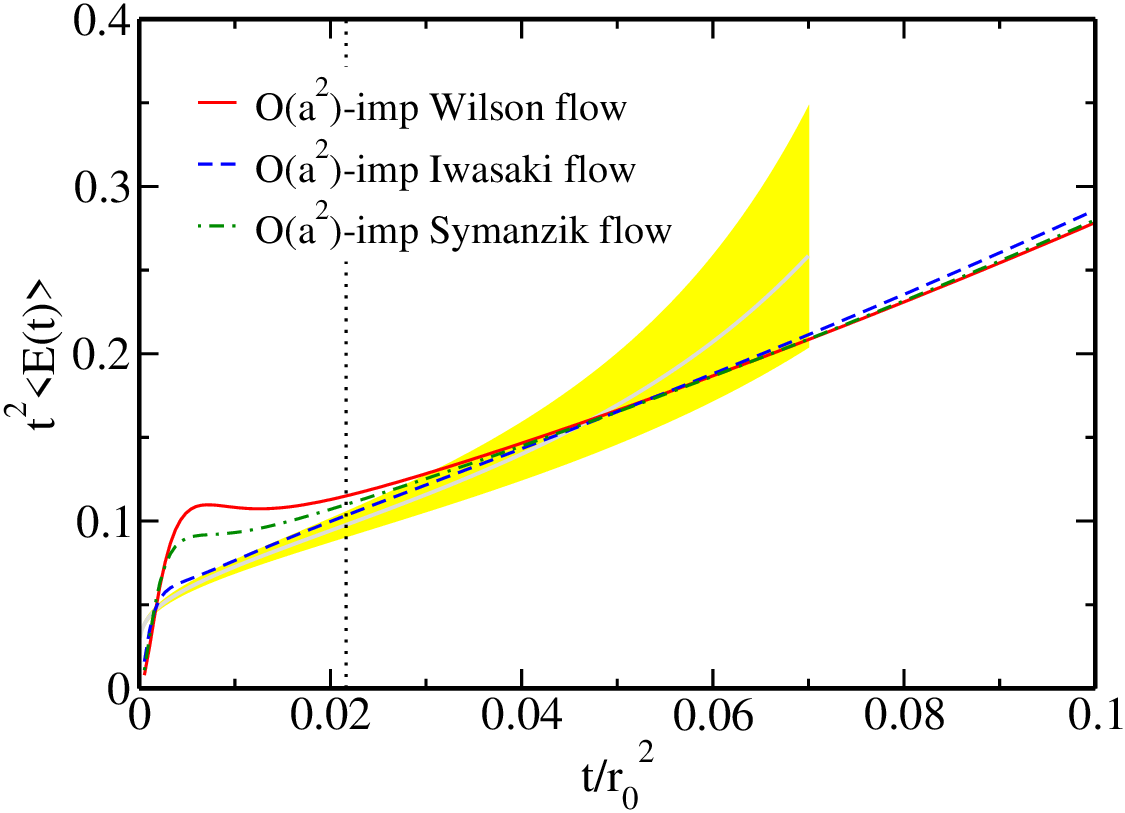}
\includegraphics[width=.32\textwidth,clip]{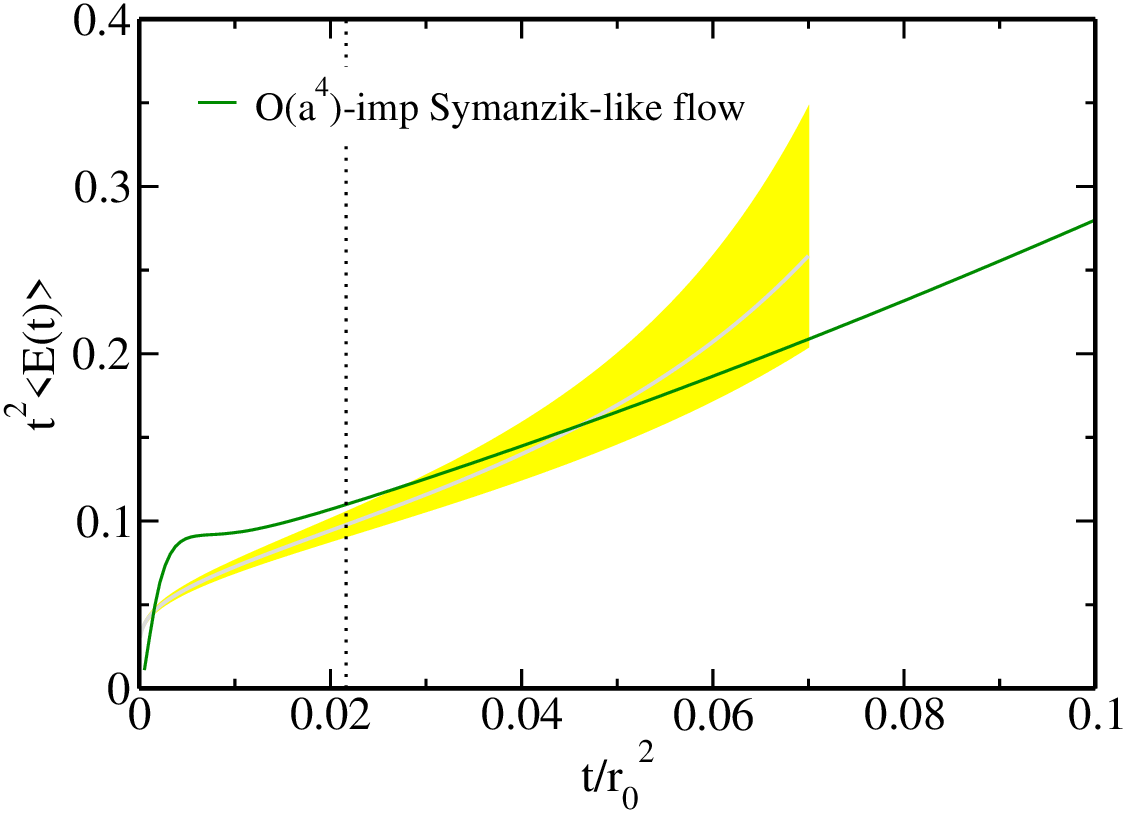}
\caption{The behavior of $t^2 \langle E (t) \rangle$ calculated on the Iwasaki gauge configurations 
at $\beta = 2.80$ as functions of $t/r_0^2$. 
The three panels show results for unimproved flows (left), $\mathcal{O}(a^2)$-imp flows (center),
and the $\mathcal{O}(a^4)$-imp Symanzik-like flow (right) (using the same graphical conventions as in Fig.~\ref{fig:beta617}.)}
 \label{fig:beta280}
 \end{figure*}
%

%
%
\begin{figure*}[ht]
\includegraphics[width=.32\textwidth,clip]{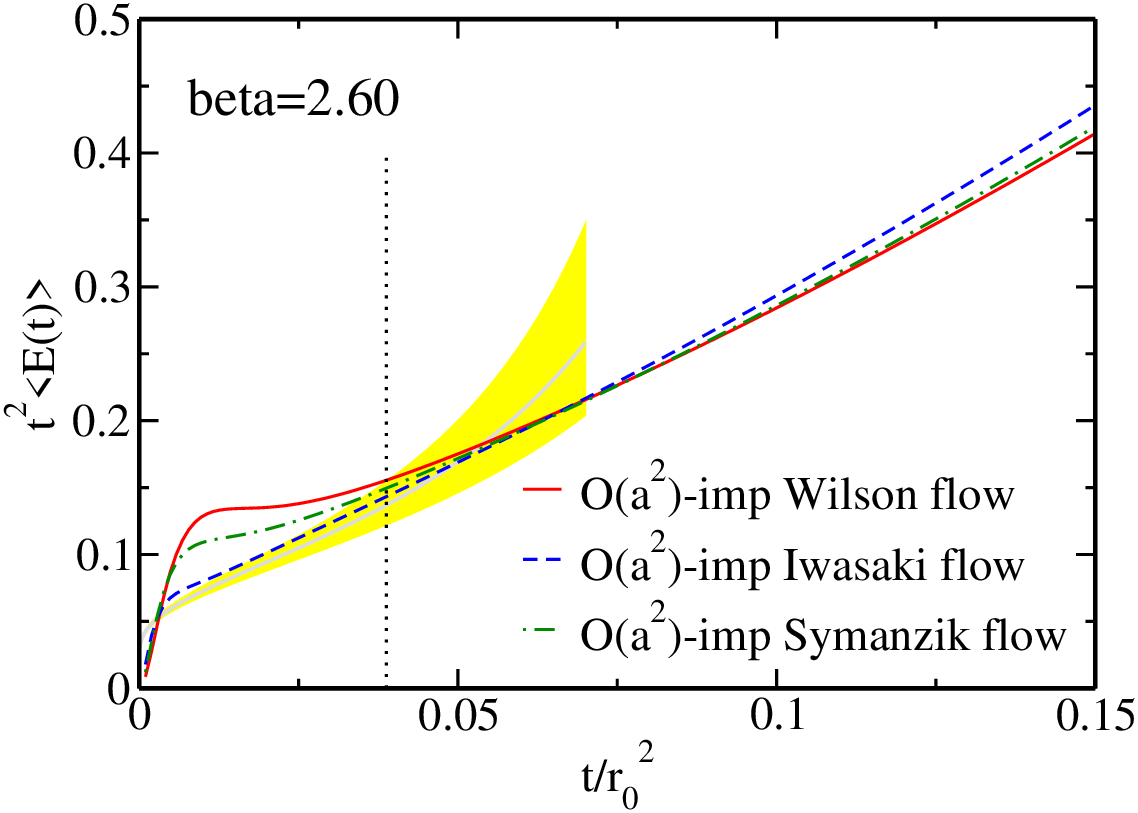}
\includegraphics[width=.32\textwidth,clip]{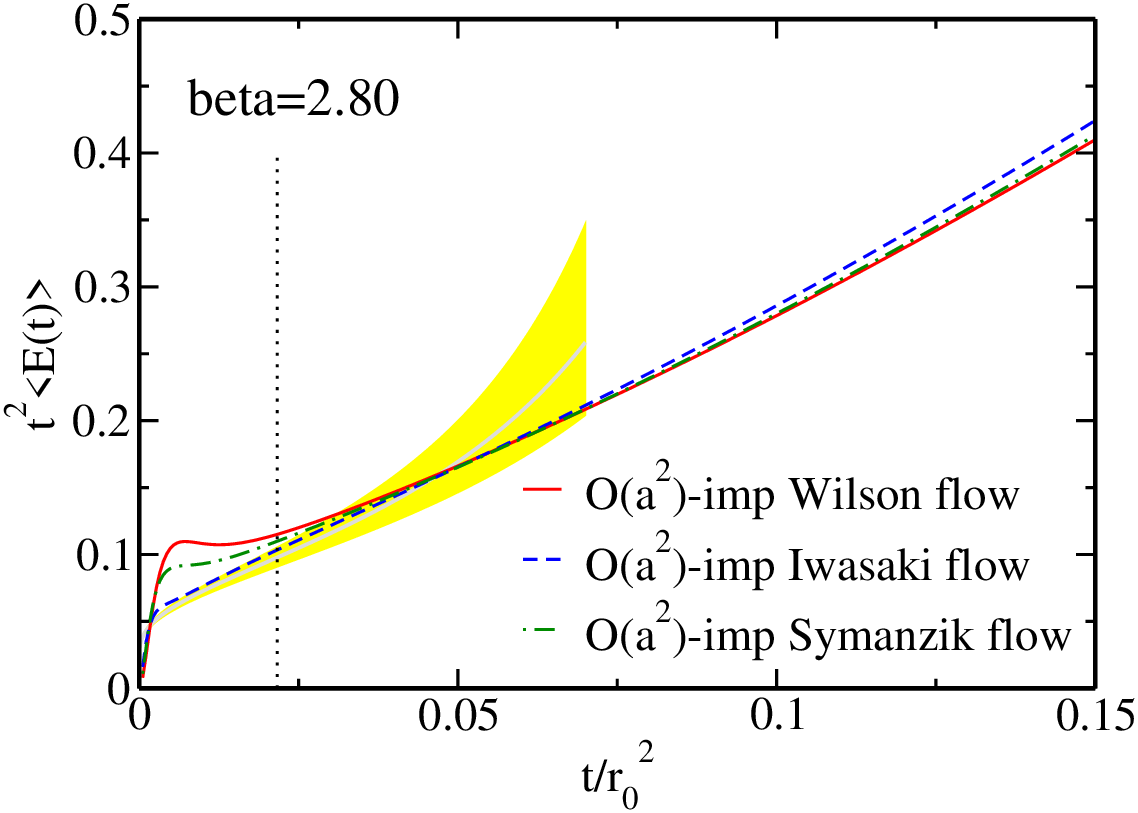}
\includegraphics[width=.32\textwidth,clip]{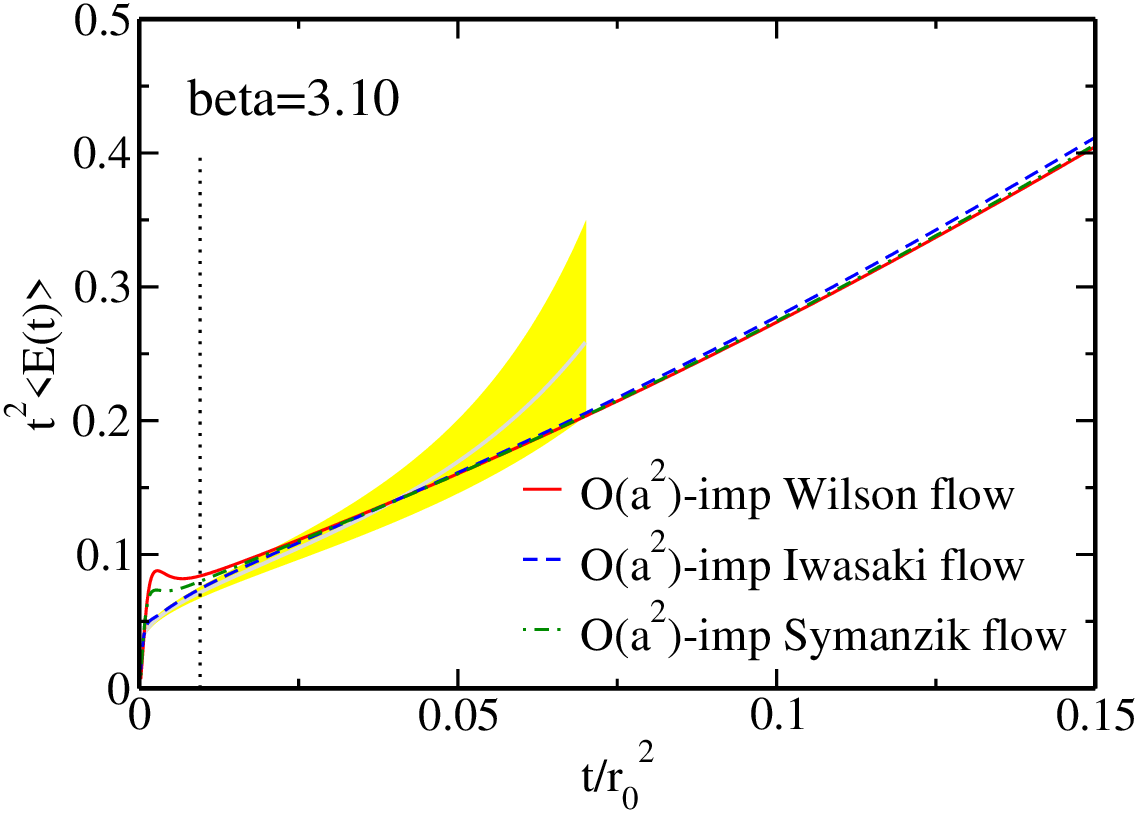}
\caption{The behavior of $t^2 \langle E (t) \rangle$ obtained 
from $\mathcal{O}(a^2)$-imp flows as functions of $t/r_0^2$ on the Iwasaki gauge configurations.
The three panels show the results calculated at $\beta=2.60$ (left), $\beta=2.80$ (center), and
$\beta=3.10$ (right) (using the same graphical conventions as in Fig.~\ref{fig:beta617}.)}
\label{fig:IG_o2imp}
\end{figure*}

Figure~\ref{fig:beta280} shows the $t$ dependence of $t^2 \langle E (t) \rangle$ calculated at $\beta = 2.80$. 
The three panels show results for unimproved flows (left), $\mathcal{O}(a^2)$-imp flows (center),
and the $\mathcal{O}(a^4)$-imp Symanzik-like flow (right).
For the unimproved case (left panel), the Wilson flow result is closest to the continuum 
perturbative calculation as same in the case of the Wilson gauge configurations.
According to the size of tree-level discretization errors of $t^2 \langle E (t) \rangle$ summarized 
in Table~\ref{tab:Tree-level lattice effects}, the three flow results move away from the continuum 
perturbative result. 

%
%
\begin{table}[hb]
\caption{
Simulation parameters of four ensembles generated by the Iwasaki gauge action (IG).
The values of the Sommer scale $r_0$ and lattice spacing $a$ are
taken from Ref.~\cite{Takeda:2004xha}. $N_{\textrm{conf}}$ is the number of gauge configurations.
\label{tab:Iset-up}
}
\begin{ruledtabular} 
\begin{tabular}{lccccc}
 $\beta$ (Action) & ($L^3 \times T$) & $r_0/a$ & $a$ [fm] & $\sim La$ [fm] & $N_{\textrm{conf}}$\cr
\hline
2.60 (IG)& $24^3\times 48$ & 5.078(64)& 0.0985(12)&  2.36&  100\cr
2.80 (IG)& $32^3\times 64$&  6.798(57)& 0.0736(6) &  2.36&  100\cr
3.10 (IG)& $48^3\times 96$&  10.23(7)& 0.0489(3) &  2.35&  100\cr
3.10 (IG)& $32^3\times 32$&  10.23(7)& 0.0489(3) &  1.56&  100\cr
\end{tabular}
\end{ruledtabular} 
\end{table}

In the center and right panels of Fig.~\ref{fig:beta280}, it is observed that 
our proposal of tree-level improvements works as well as the case of 
the Wilson gauge configurations (see Fig.~\ref{fig:beta617} for comparison).
Here we note that we omit the result obtained from another $\mathcal{O}(a^4)$-imp flow (namely,
the ``positive rectangle flow") in the right panel of Fig.~\ref{fig:beta280}.
This is simply because the positive rectangle flow yields a negative value of 
$\langle E (t) \rangle$ in the entire positive flow time region ($t>0$)~\cite{rectangle}. 
The numerical results for the $\mathcal{O}(a^2)$-imp Symanzik flow (center panel)
and the $\mathcal{O}(a^4)$-imp Symanzik-like flow (right panel) 
mostly coincide since the $C_4$ coefficient of the $\mathcal{O}(a^2)$-imp Symanzik flow
is tiny, as shown in Table~\ref{tab:Tree-level lattice effects}. 
Indeed, the rectangle coefficient ($c_f = - 0.0838$) of the $\mathcal{O}(a^4)$-imp Symanzik-like 
flow is quite close to the Symanzik action ($c_f = -1/12 \approx -0.08333$).
There is a neither qualitative nor quantitative difference between the results for 
the $\mathcal{O}(a^2)$-imp Symanzik and $\mathcal{O}(a^4)$-imp Symanzik-like flows.
For these reasons, we hereafter focus on the $\mathcal{O}(a^2)$-imp flows.

Let us take a closer look at the results for $\mathcal{O}(a^2)$-imp flows (center panel). 
Among the three types of $\mathcal{O}(a^2)$-imp flows, 
the $\mathcal{O}(a^2)$-imp Iwasaki flow is closest to the continuum 
perturbation calculation, while the other flows largely overshoot the continuum counterpart 
in the relatively small-$t$ regime ($0.005 \alt t/r_0^2 \alt 0.02$).
This is not observed in the case of the Wilson gauge configurations, where
the three types of $\mathcal{O}(a^2)$-imp flow mostly coincide
near the continuum counterpart even in the small-$t$ regime up to $t/r_0^2 \approx 0.01$.
However, it should be remembered that the lattice-spacing dependence 
of the tree-level contribution is classified by powers of $a^2/t$ 
as defined in Eq.~(\ref{eq:flowed_energy_lattice}).
In the strict sense, the tree-level improvement program 
proposed by Fodor {\it et al.}~\cite{Fodor:2014cpa}, where
the tree-level contributions of $t^2\langle E(t) \rangle$ are classified by powers of $a^2/t$,
is supposed to be valid only in the region of $t/r_0^2 \gg (a/r_0)^2\approx 0.02$
at $\beta=2.80$ (Iwasaki) or $\beta=6.17$ (Wilson).
 
In the large-$t$ regime ($t/r_0^2 > 0.05$), the differences among the results from the three 
$\mathcal{O}(a^2)$-imp flows gradually appear and also increase for a larger value of $t$.
As explained in Sec.~\ref{Sec:3-A}, the origin of these differences is 
the non-negligible $\mathcal{O}(g^{2n} a^2)$ corrections beyond the 
tree-level discretization effects. To see this point, we show 
the results from the three $\mathcal{O}(a^2)$-imp flows at three different lattice spacings 
in Fig.~\ref{fig:IG_o2imp}. The left panel is for $\beta=2.60$, the center one is 
for $\beta=2.80$, and the left one is for $\beta=3.10$. It is clear that
the differences between these three $\mathcal{O}(a^2)$-imp flows are diminished 
as we move from a coarser lattice spacing (left panel) to a finer lattice spacing (right panel). 

\begin{figure}[ht]
\includegraphics[width=.40\textwidth,clip]{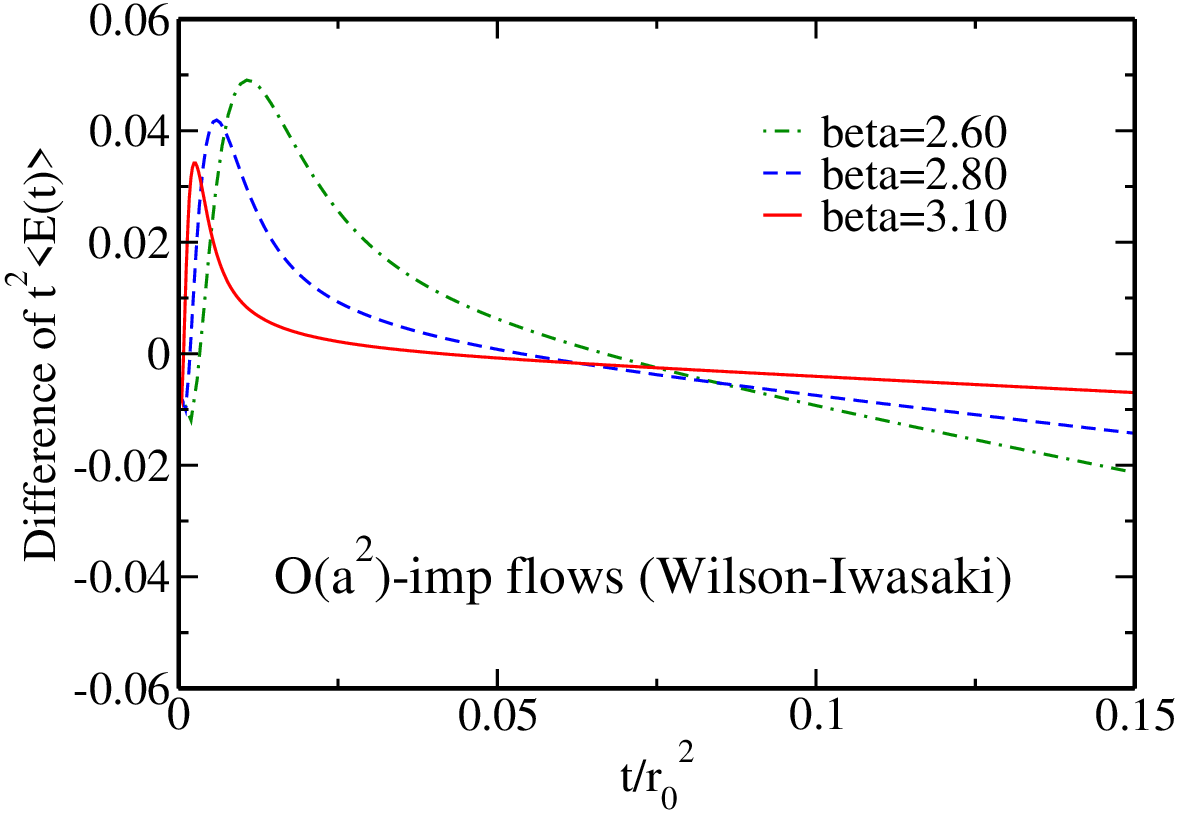}
\includegraphics[width=.40\textwidth,clip]{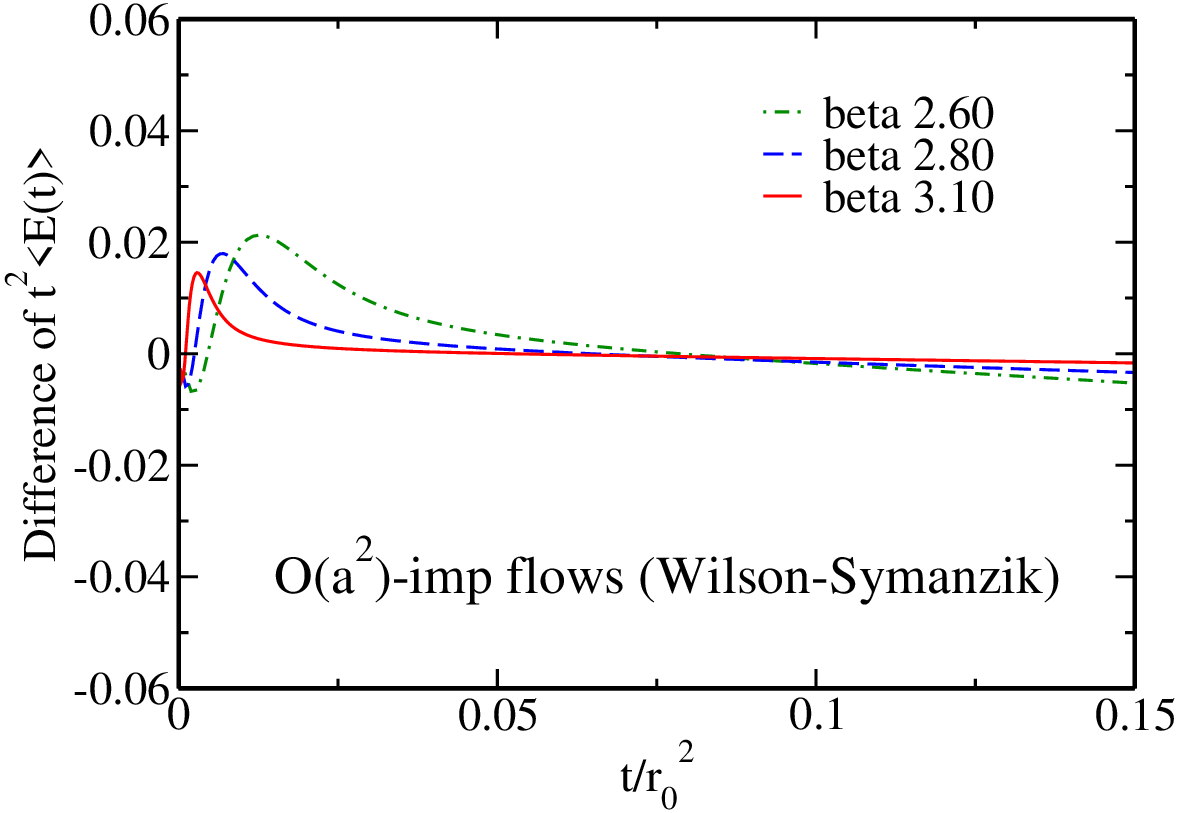}
\caption{Difference in the value of $t^2 \langle E (t) \rangle$ 
between two of the three $\mathcal{O}(a^2)$-imp flows as functions of $t/r_0^2$.
The upper (lower) panel shows the difference between $\mathcal{O}(a^2)$-imp Wilson and 
$\mathcal{O}(a^2)$-imp Iwasaki (Symanzik) flows.
Green dot-dashed, blue dashed, and red solid curves represent results 
at $\beta=2.60$, 2.80, and 3.10.
}
\label{fig:IG_t2Edif}
\end{figure}

\begin{figure}[bt]
\includegraphics[width=.40\textwidth,clip]{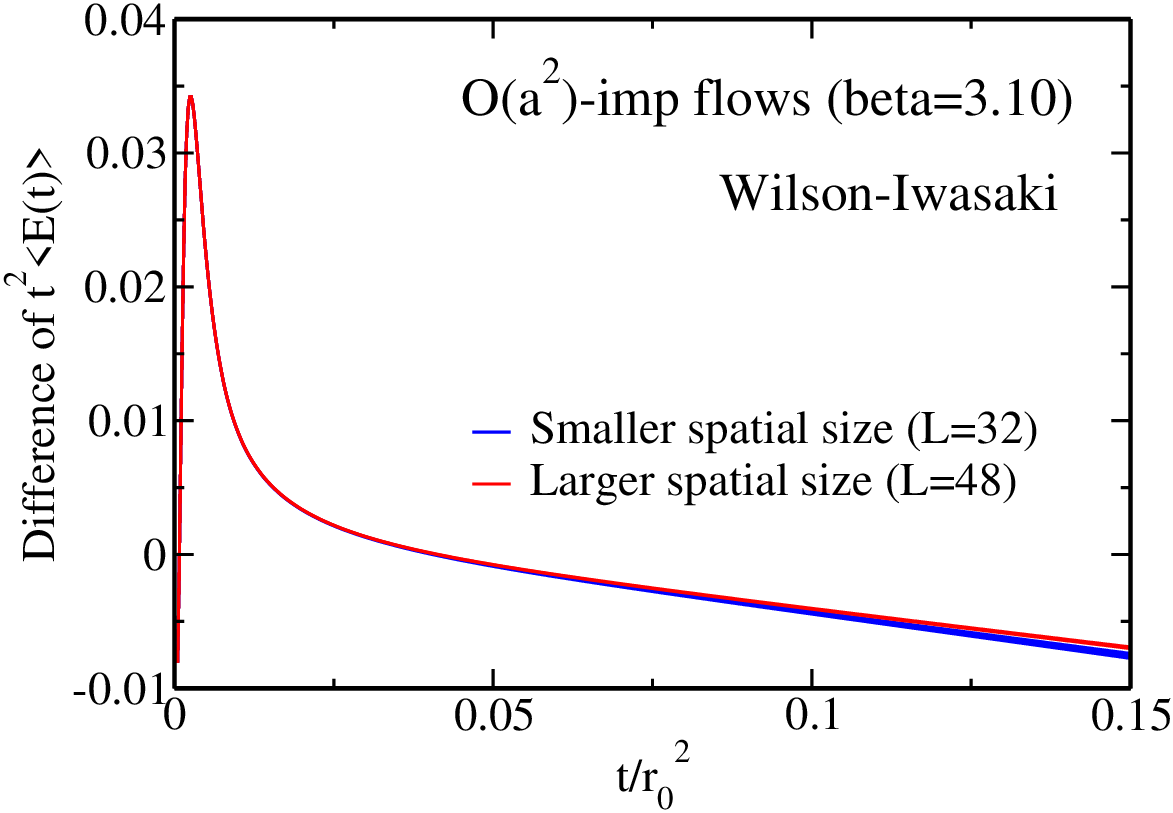}
\includegraphics[width=.40\textwidth,clip]{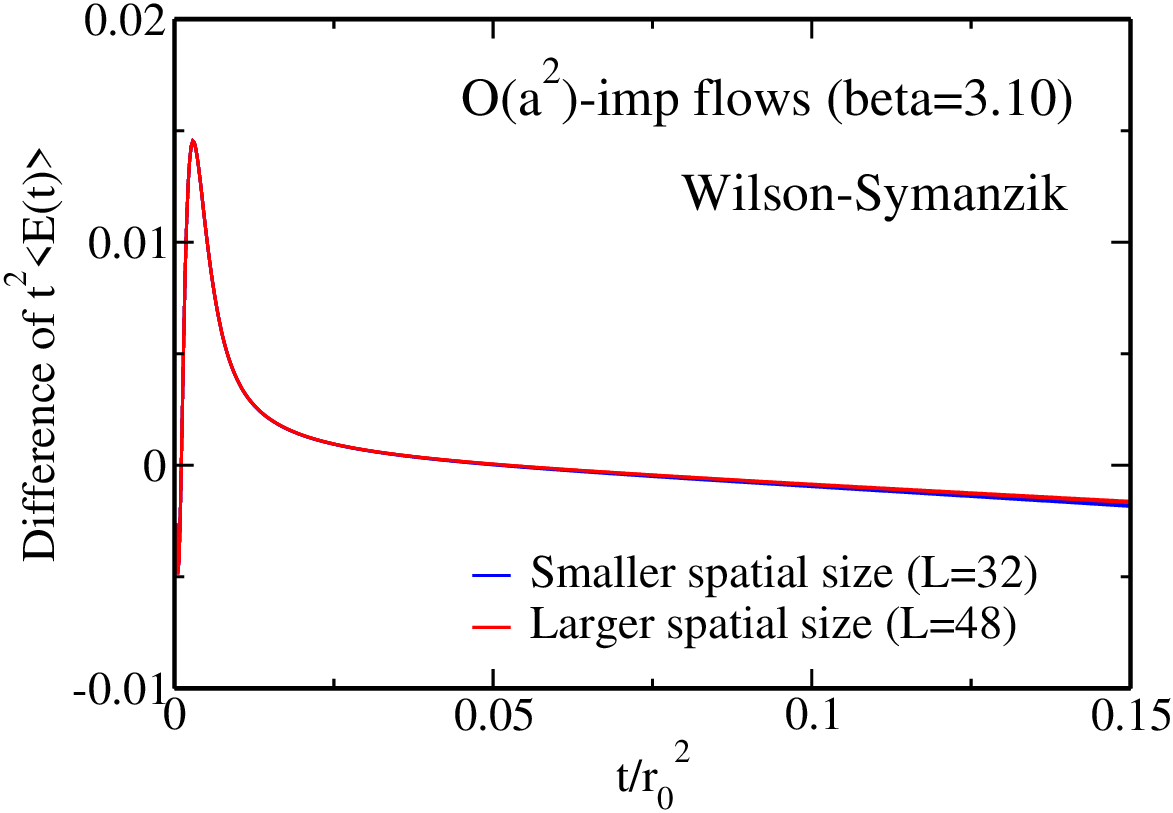}
\caption{Finite-volume dependence in differences in $t^2 \langle E (t) \rangle$
between two of the three $\mathcal{O}(a^2)$-imp flows as functions of $t/r_0^2$. The upper (lower) panel 
shows the difference between $\mathcal{O}(a^2)$-imp Wilson and 
$\mathcal{O}(a^2)$-imp Iwasaki (Symanzik) flows (using the same graphical conventions as in Fig.~\ref{fig:WG_t2Edif_FV}.)
}
\label{fig:IG_FV}
\end{figure}

In Fig.~\ref{fig:IG_t2Edif} we also plot the differences in the values
of $t^2 \langle E (t) \rangle$ between $\mathcal{O}(a^2)$-imp flows 
as functions of $t/r_0^2$. 
The upper (lower) panel shows the difference between $\mathcal{O}(a^2)$-imp Wilson 
and $\mathcal{O}(a^2)$-imp Iwasaki (Symanzik) flows.
Green dot-dashed, blue dashed, and red solid curves represent results at $\beta=2.60$,
2.80, and 3.10. These figures clearly show
that the differences appearing in both the smaller $t$ region ($t/r_0^2 \alt 0.05$) and 
larger $t$ region ($t/r_0^2 > 0.05$) stem from some discretization errors.
Through a direct comparison of the results obtained in two different lattice volumes 
($48^3\times 96$ and $32^3\times 32$) at the finer lattice spacing ($\beta=3.10$), we
confirm that the finite-volume effects in calculations using the Iwasaki gauge configurations are 
also negligible in the range of $t/r_0^2 \alt 0.15$, as depicted in Fig.~\ref{fig:IG_FV}.

%
%
%

%
%
%

\section{Scaling behavior and continuum limit of action density} 
\label{Sec:4}
In this section, 
to see how much improvement we get from our proposal, 
we first show the behavior of ${\sqrt{t_X}}/{r_0}$ as a function of $a^2/t_X$.
A similar plot was used in Ref.~\cite{Luscher:2010iy} for a discussion of the finite-lattice-spacing effects.
Here, $r_0$ and $t_X$ are the Sommer scale~\cite{Sommer:1993ce} and 
the new reference scale defined in Eq.~(\ref{eq:new_scale}), respectively.
Although we would like to discuss the discretization errors solely on the reference scale $t_X$, 
the ratio ${\sqrt{t_X}}/{r_0}$ also contains the finite-lattice-spacing effects on the Sommer scale,
which is determined at finite lattice spacing. 

For this reason, instead of ${\sqrt{t_X}}/{r_0}$, we will later discuss the scaling 
behavior of the dimensionless combination of $\sqrt{t_X}\Lambda_{\overline{\rm MS}}$ 
with the QCD parameter $\Lambda$ 
in the $\overline{\rm MS}$ scheme following the analysis of Ref.~\cite{Gockeler:2005rv}.
The parameter $\Lambda_{\overline{\rm MS}}$ is determined through a matching between 
the lattice bare coupling $g^2_0$ to the $\overline{\rm MS}$ running coupling $g^2$
with the help of perturbation theory. There is, at least, no power-like
dependence on the lattice artifacts 
in the determination of $\Lambda_{\overline{\rm MS}}$.
On the other hand, 
the choice of the smaller $X$---which
gives the higher energy scale $\sim 1/\sqrt{t_X}$---is desirable to ensure the applicability of 
a perturbative matching procedure.

In the previous section, we have found that 
the lattice discretization errors of the energy density $\langle E \rangle$  
around $t/r_0^2\approx 0.05$ are well controlled by the $\mathcal{O}(a^2)$ or
$\mathcal{O}(a^4)$-imp flows. The corresponding $X$ is roughly 0.15, which is 
a factor of 2 reduction from the original choice of $X=0.3$. In this study, we will 
later evaluate $t_X$ for $X=0.15$ and 0.3 
as typical examples.


\subsection{The behavior of ${\sqrt{t_X}}/{r_0}$}
\label{Sec:4-L}

In Ref.~\cite{Luscher:2010iy}, the value of ${\sqrt{t_{0.3}}}/{r_0}$ was evaluated
from both the plaquette- and clover-type energy densities by using the Wilson flow. 
We here show the behavior of ${\sqrt{t_X}}/{r_0}$ as a function of $a^2/t_X$
for several combinations of two gauge actions (WG and IG) and various flows
in order to demonstrate the feasibility of our tree-level  improvement proposal.

\begin{figure*}[ht]
\includegraphics[width=.40\textwidth,clip]{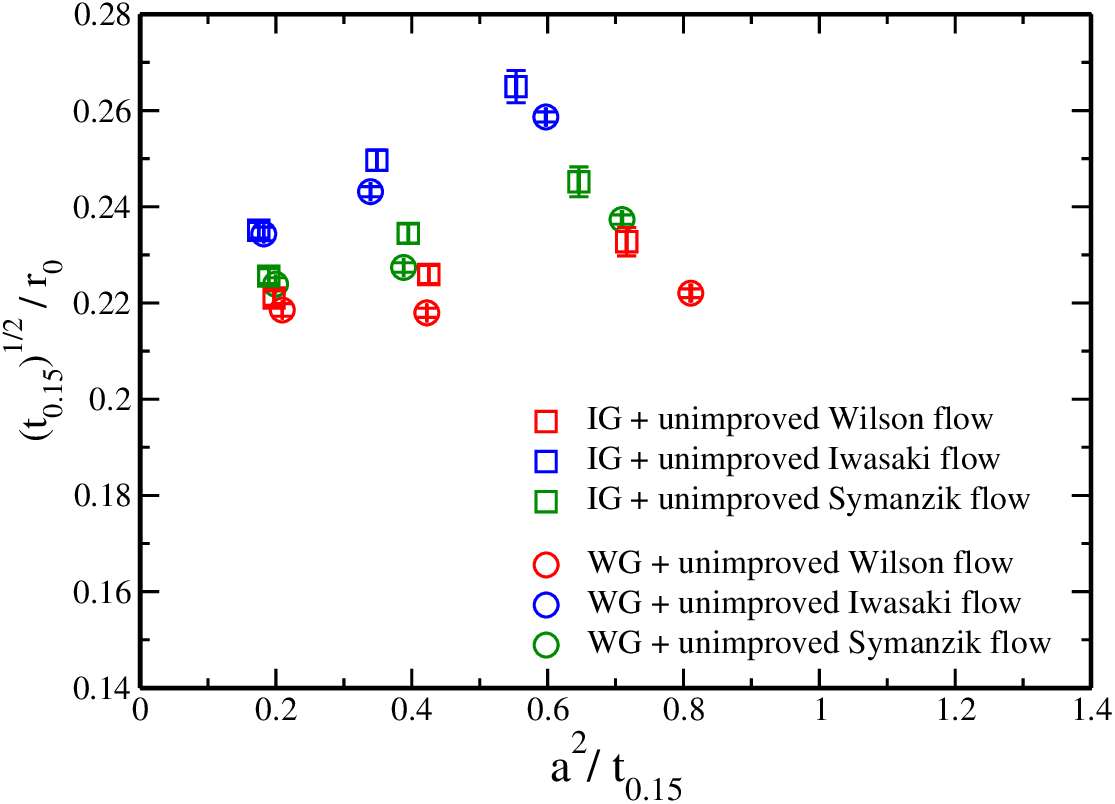}
\includegraphics[width=.40\textwidth,clip]{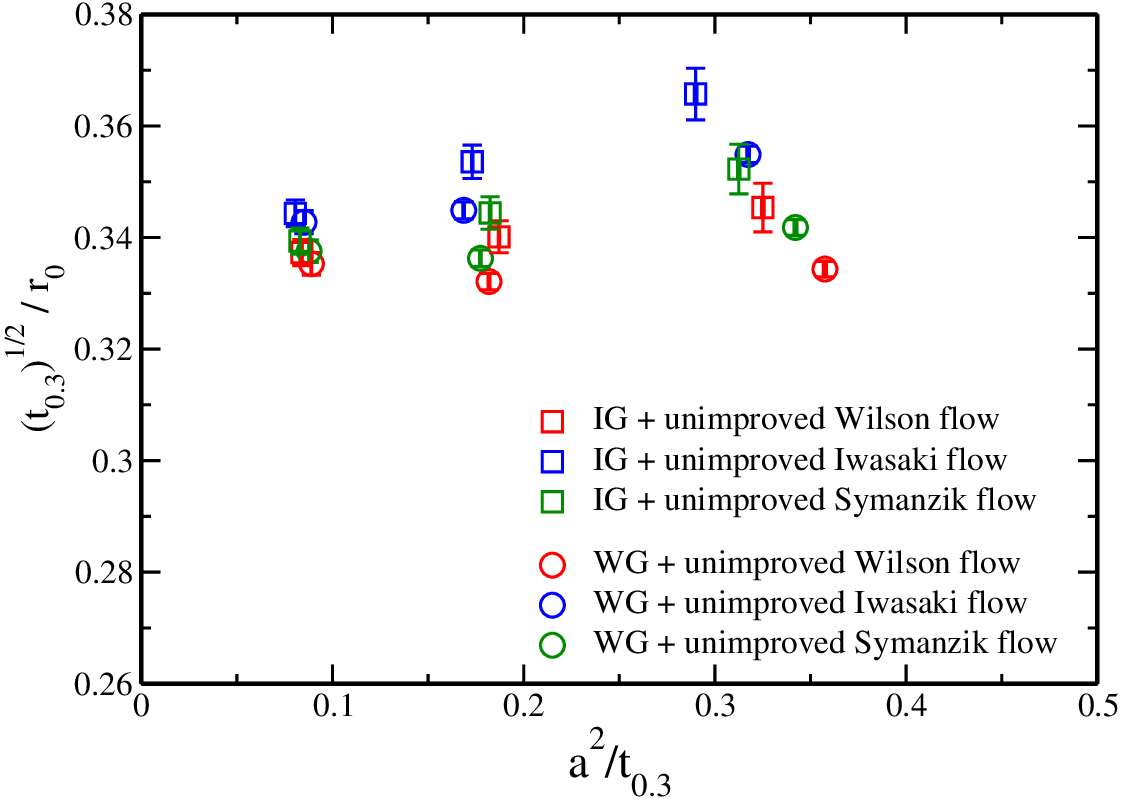}
\includegraphics[width=.40\textwidth,clip]{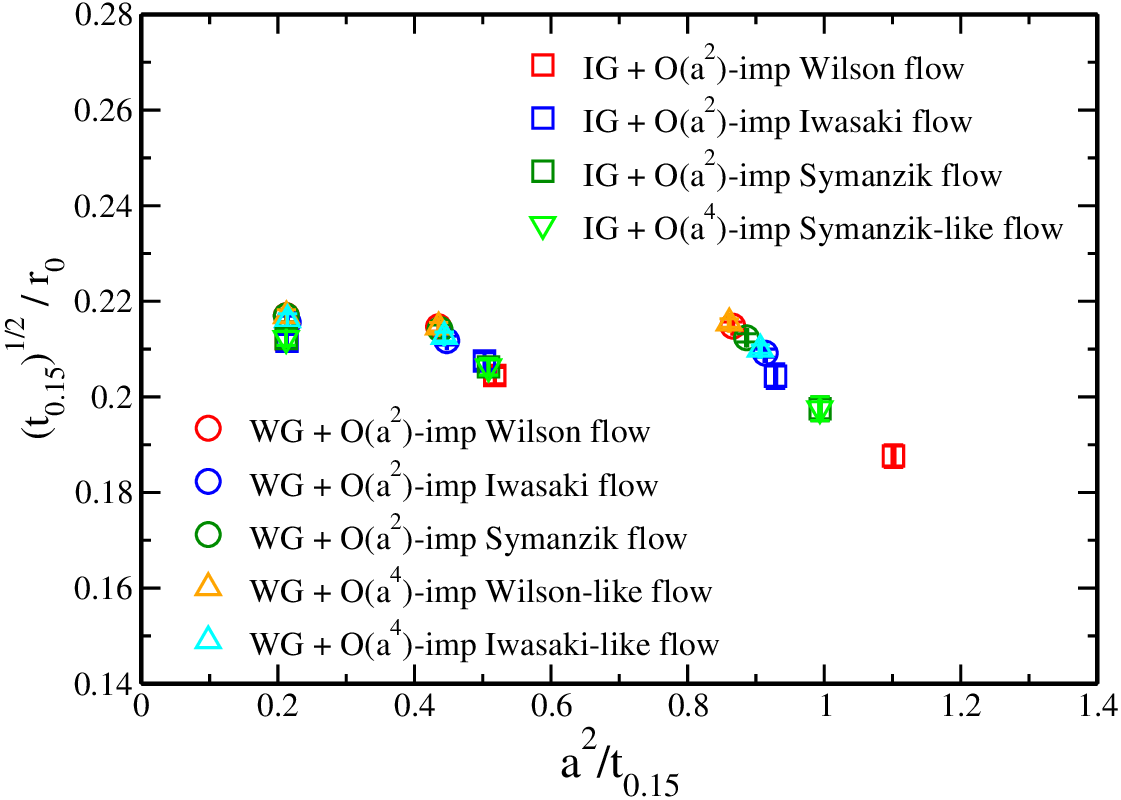}
\includegraphics[width=.40\textwidth,clip]{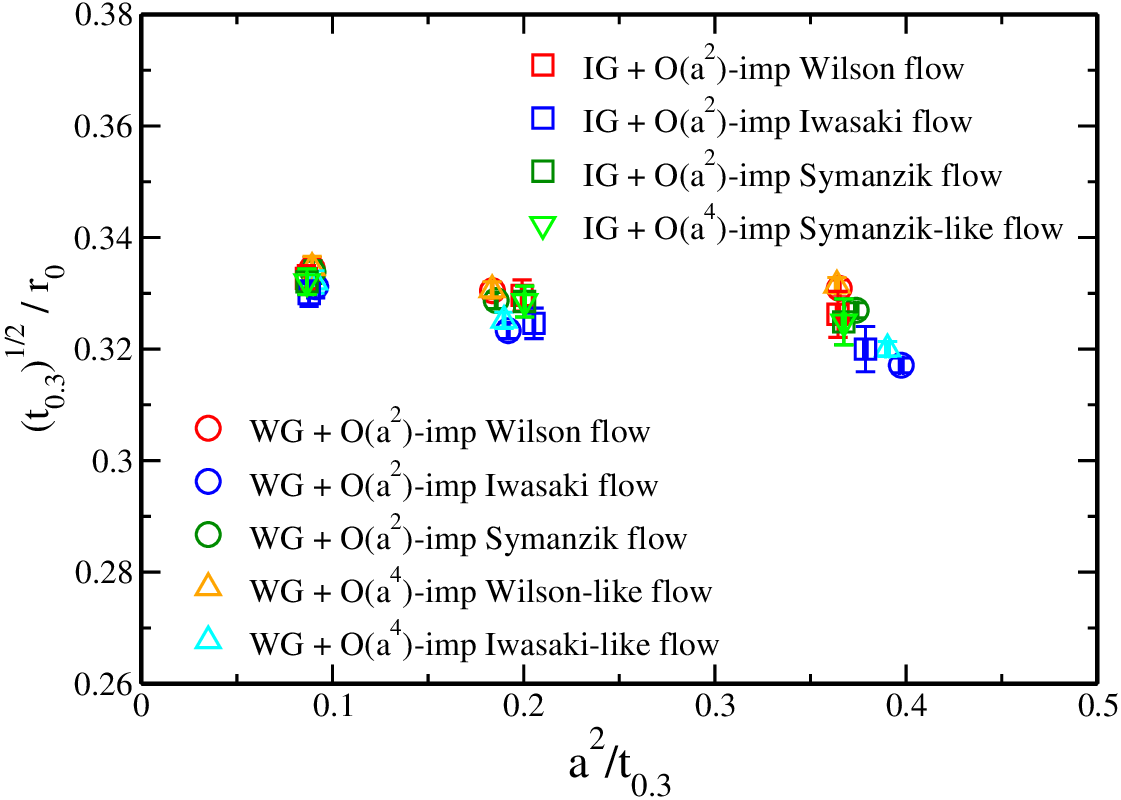}
\caption{The scaling behaviors of $\sqrt{t_X}/r_0$
versus $a^2/t_X$ for choices of $X=0.15$ (left panels) and $X=0.3$ (right panels). 
The two upper panels show the results obtained by unimproved flows, while the two lower panels are for
the results obtained by $\mathcal{O}(a^2)$ and $\mathcal{O}(a^4)$-imp flows.
In the legends of these figures, WG (IG) stands for the Wilson (Iwasaki) gauge action
used in generating gauge configurations. The values of $r_0/a$
are taken from Ref.~\cite{Guagnelli:1998ud} for WG and Ref.~\cite{Takeda:2004xha} for IG,
as summarized in Tables~\ref{tab:Wset-up} and \ref{tab:Iset-up}.
}
\label{fig:sommer}
\end{figure*}

To evaluate the ratio ${\sqrt{t_X}}/{r_0}$, we use the values of 
$r_0/a$ determined in Refs.~\cite{{Guagnelli:1998ud},{Takeda:2004xha}}. 
Figure~\ref{fig:sommer} just shows a feature of how much improvement has been achieved 
by tree-level $\mathcal{O}(a^2)$ and $\mathcal{O}(a^4)$-imp flows.
The results for ${\sqrt{t_X}}/{r_0}$ are plotted against $a^2/t_X$ for 
various flows carried out on two types of gauge configurations (WG and IG).
The two upper panels show the results obtained from unimproved flows, while 
the results for $\mathcal{O}(a^2)$ and $\mathcal{O}(a^4)$-imp flows
are presented in the two lower panels.

For the unimproved flows (upper panels), the results for both cases of $X=0.15$ (left)
and 0.3 (right) very much depend on the choice of the gauge action and the flow. 
However, once any tree-level improvement is achieved, the differences among 
the different choices of the gauge action and the flow become significantly 
diminished. 
If one takes a closer look at the figures in the two lower panels of Fig.~\ref{fig:sommer}, 
the results calculated on the same gauge action (WG or IG) seem to be clustered while
the gauge action dependence remains visible. However, we consider that 
the gauge action dependence would be partly attributed to the systematic uncertainties 
due to the scale setting by the Sommer scale, which was not directly 
determined in our numerical simulations. 
Furthermore, as described previously, the value of $r_0$ (which is determined in lattice QCD) 
itself receives the lattice discretization errors independently.
Therefore, the ratio ${\sqrt{t_X}}/{r_0}$
is not an appropriate quantity to discuss the lattice artifacts solely on  
$t_X$. Instead of the Sommer scale, we will then use $\Lambda_{\overline{\rm MS}}$
for the scale setting in the next section.

%
%

\subsection{Conversion to the $\overline{\rm MS}$ scheme}
\label{Sec:4-A}
The parameter $\Lambda_{\overline{\rm MS}}$ is determined through a matching between 
the lattice bare coupling $g^2_0$ to the $\overline{\rm MS}$ running coupling $g^2$
with help of perturbation theory. However, it is well known that the lattice perturbative 
expansions are poorly convergent. 
We thus introduce the tadpole-improved (TI) coupling, which is defined by
%
%
\begin{equation}
g_{\rm TI}^2(a)=\frac{g_0^2}{u_0^4}=\frac{6}{\beta u_0^4}
\end{equation}
with 
%
%
\begin{equation}
u_0^4=(1-8c_{\rm rect})P+8c_{\rm rect} R
\end{equation}
where $P$ and $R$ represent the expectation values of the the path-ordered plaquette 
and rectangle products of link variables, respectively \cite{Gockeler:2005rv,AliKhan:2001wr}.
The tadpole-improved coupling can {\it boost} the slow convergence of a power series 
in the lattice bare coupling. 

In order to evaluate the $\Lambda_{\overline{\rm MS}}$, 
let us consider a conversion from the boosted lattice scheme to 
the $\overline{\textrm{MS}}$ scheme as follows. 
The running coupling in the $\overline{\textrm{MS}}$ scheme,
$g^2$, is given by the following formula 
as a power series in the boosted coupling $g_{\rm TI}^2(a)$, 
up to ${\cal O}(g_{\rm TI}^4)$:
%
%
\begin{eqnarray}
\label{eq:gauge_coupling}
\frac{1}{g^2(\mu)} &=& \frac{1}{g^2_{\rm TI}(a)}
+2b_0\ln(\mu a) - t_1^{\rm TI} \nonumber\\
&&+\left(2b_1\ln(\mu a) - t_2^{\rm TI} \right) g^2_{\rm TI}(a)+{\cal O}(g^4_{\rm TI}),
\end{eqnarray}
where $t_1^{\rm TI}$ and $t_2^{\rm TI}$ denote the one-loop and two-loop conversion variables
with the first two coefficients of the $\beta$ function $b_0=\frac{11}{(4\pi)^2}$ 
and $b_1= \frac{102}{(4\pi)^4}$ being the universal coefficients 
in the pure Yang-Mills theory~\cite{Gockeler:2005rv}. This conversion formula is
fully determined by the NLO perturbation theory. However, the two-loop conversion 
variable, $t_2^{\rm TI}$, is not known for 
the case of $c_{\rm rect}=-0.331$, since the three-loop term 
of the lattice $\beta$ function is not available 
for the RG-improved gauge action~\cite{Skouroupathis:2007mq}.
For the standard Wilson ($c_{\rm rect}=0$) and the Iwasaki ($c_{\rm rect}=-0.331$)
gauge actions, the currently known results for conversion variables
in the tadpole-improved lattice scheme are summarized in Table~\ref{tab:TIcoeff}.

We next choose the renormalization scale $\mu$ to remove the $\mathcal{O}(g_{\rm TI}^0)$ 
coefficient in Eq.~(\ref{eq:gauge_coupling}) for its rapid convergence. 
To achieve this, the scale $\mu$ is set to 
%
%
\begin{equation}
\mu=\mu_{\ast}=\frac{1}{a}\exp \left(\frac{t_1^{\rm TI}}{2b_0}\right).
\end{equation}
In this choice, the explicit $a$ dependence 
appears only at the level of $\mathcal{O}(g^4_{\rm TI})$ in Eq.~(\ref{eq:gauge_coupling})
with two-loop conversion variables. Therefore, the lattice discretization errors on 
the determination of $\Lambda_{\overline{\rm MS}}$ become negligible 
in the weaker coupling region. 
This scale choice is called ``method I" in Ref.~\cite{Gockeler:2005rv}
and reduces Eq.~(\ref{eq:gauge_coupling}) to
%
%
\begin{equation}
\frac{1}{g^2(\mu_{\ast})} 
=\left\{
\begin{array}{ll}
 \frac{1}{g^2_{\rm TI}(a)} & (\textrm{one-loop}) 
 \cr
 \frac{1}{g^2_{\rm TI}(a)}+\left(\frac{b_1}{b_0}t_1^{\rm TI} - t_2^{\rm TI} \right) g^2_{\rm TI}(a)
 & (\textrm{two-loop})
\end{array}\right.
\label{eq:matching_formula}
\end{equation}
which correspond to conversions from the boosted coupling to the $\overline{\textrm{MS}}$
coupling at one-loop (first line) and two-loop (second line) order of perturbation theory.
To evaluate $g^2(\mu_{\ast})$, we need to compute 
$P$ and $R$ numerically. We summarize our results for $P$ and $R$ in Table~\ref{tab:PandR}.

In this paper, we use the knowledge of the $\beta$ function
at three-loop order for the evaluation of the $\Lambda_{\overline{\textrm{MS}}}$ parameter
with a given $\overline{\textrm{MS}}$ coupling~\cite{Gockeler:2005rv}.
For the $\overline{\textrm{MS}}$ coupling at $\mu=\mu_{\ast}$, we use the
following formula for the $\Lambda_{\overline{\textrm{MS}}}$ parameter~\cite{Gockeler:2005rv}:
%
%
\begin{eqnarray}
\Lambda_{\overline{\textrm{MS}}} = \frac{1}{a}\exp{\biggr(\frac{t^{\rm TI}_1}{2b_0}-\frac{1}{
2b_0g^2(\mu_{\ast})}\biggl)}\left(b_0 g^2 
(\mu_{\ast})\right)^{-\frac{b_1}{2b_0^2}}\nonumber\\ \times \biggl( 1 + \frac{A}{2b_0}g^2(\mu_{\ast})\biggr)^{-p_A}\biggl( 1 + \frac{B}{2b_0}g^2(\mu_{\ast})\biggr)^{-p_B}
\label{eq:MSbar_para}
\end{eqnarray}
with
\begin{eqnarray}
A &=& b_1+\sqrt{b_1^2-4b_0b_2}, \quad p_A = -\frac{b_1}{4b^2_0} - \frac{b^2_1-2b_0b_2}{4b^2_0\sqrt{b^2_1-4b_0b_2}},\nonumber\\
B &=& b_1-\sqrt{b_1^2-4b_0b_2}, \quad p_B = -\frac{b_1}{4b^2_0} + \frac{b^2_1-2b_0b_2}{4b^2_0\sqrt{b^2_1-4b_0b_2}},\nonumber
\end{eqnarray}
where $b_2 = \frac{1}{(4\pi)^6}\frac{2857}{2}$ is the third coefficient of the $\beta$ function 
in the $\overline{\textrm{MS}}$ scheme\cite{correction}. 
We thus can determine the $\Lambda_{\overline{\textrm{MS}}}$ parameter through 
Eq.(\ref{eq:MSbar_para}) by using the value of $g^2(\mu_{\ast})$
evaluated from either the one-loop or two-loop conversion formula in Eq.~(\ref{eq:matching_formula}).
The dimensionless combination of $\sqrt{t_X}\Lambda_{\overline{\textrm{MS}}}$ is finally 
obtained together with the numerically computed value of $t_X/a^2$.
If we stress that the one-loop (two-loop) formula in Eq.~(\ref{eq:matching_formula}) is used for the conversion between two schemes, 
the resulting $\Lambda$ parameter in the ${\overline{\textrm{MS}}}$ scheme
from Eq.(\ref{eq:MSbar_para}) is denoted by $\Lambda^{\text{one-loop}}_{\overline{\rm MS}}$
($\Lambda^{\text{two-loop}}_{\overline{\rm MS}}$).

%
%
\begin{figure*}[ht]
\includegraphics[width=.40\textwidth,clip]{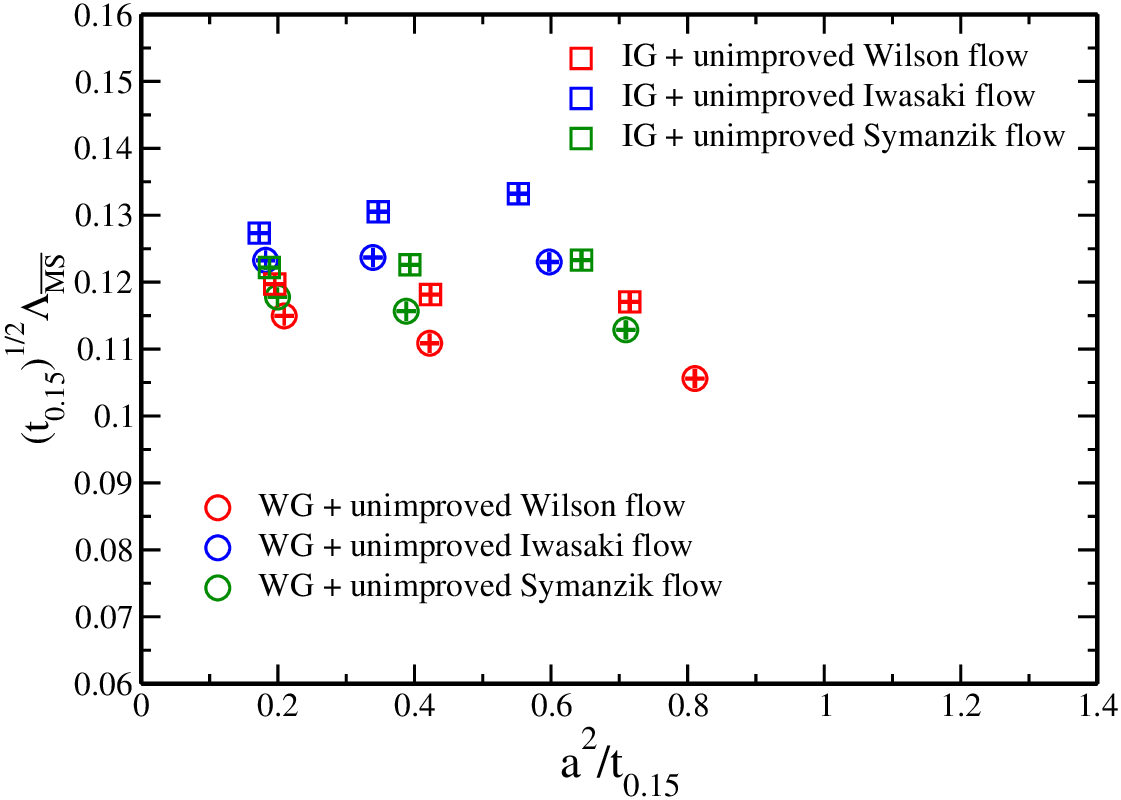}
\includegraphics[width=.40\textwidth,clip]{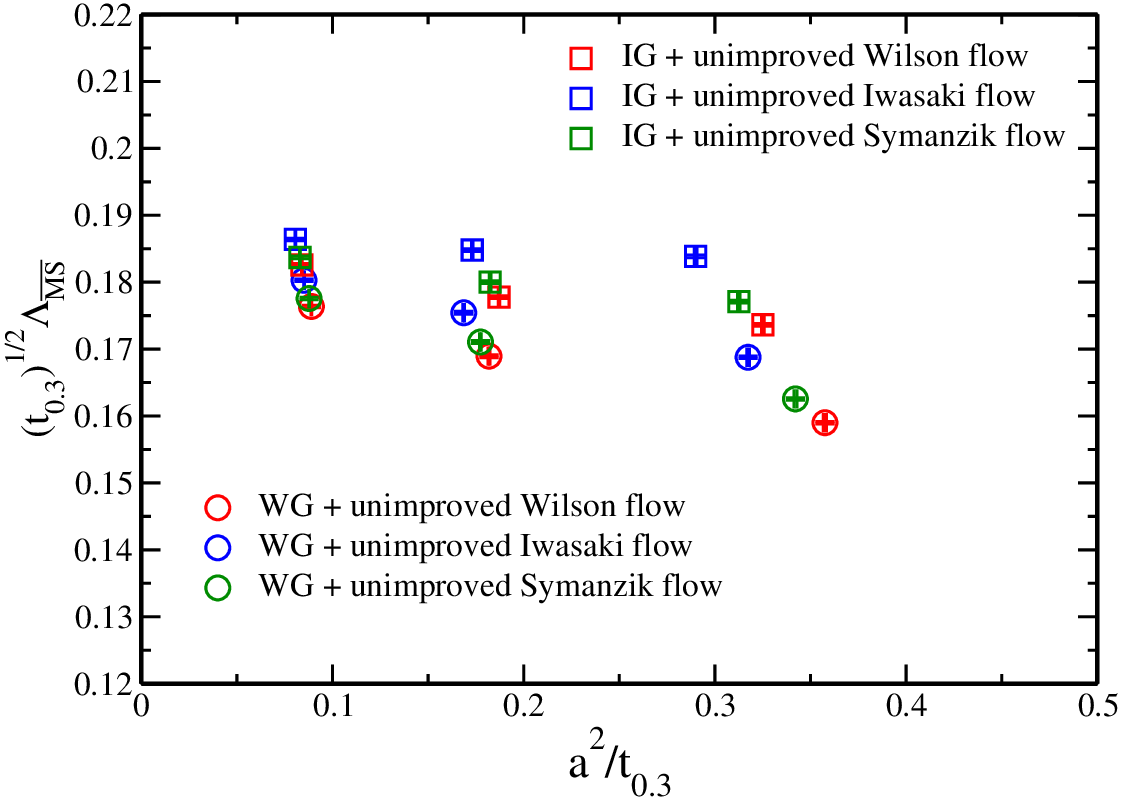}
\includegraphics[width=.40\textwidth,clip]{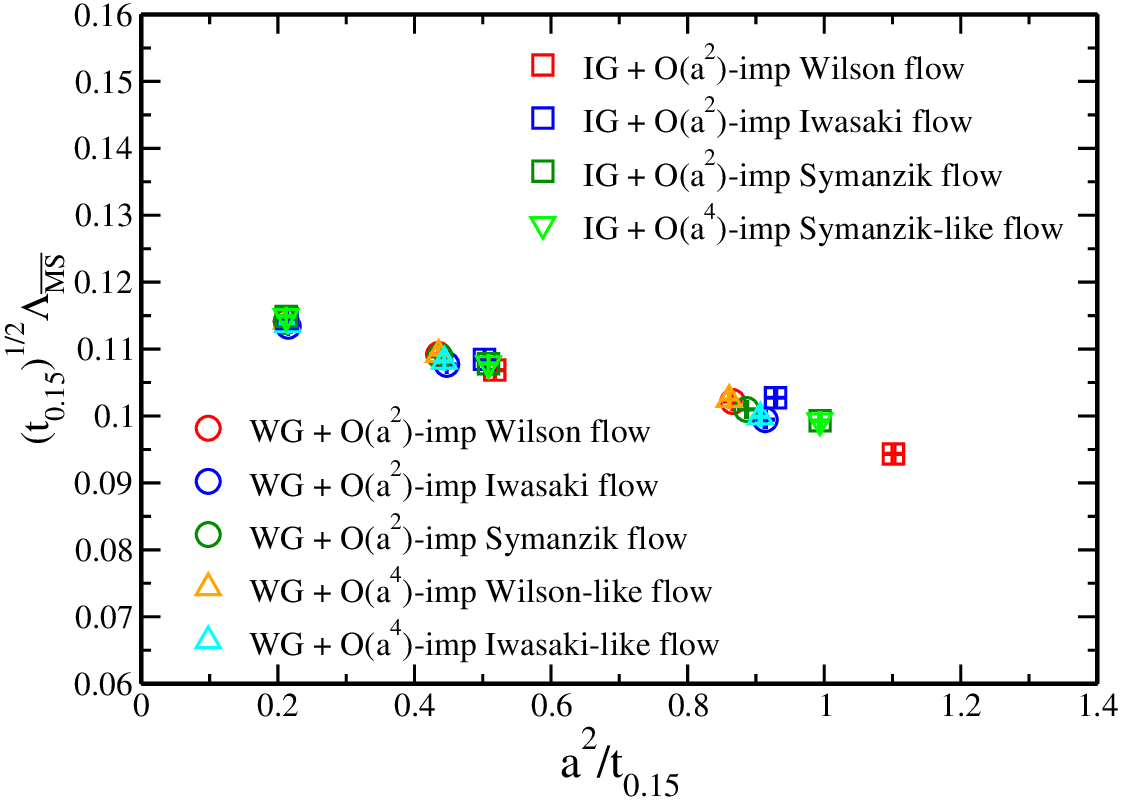}
\includegraphics[width=.40\textwidth,clip]{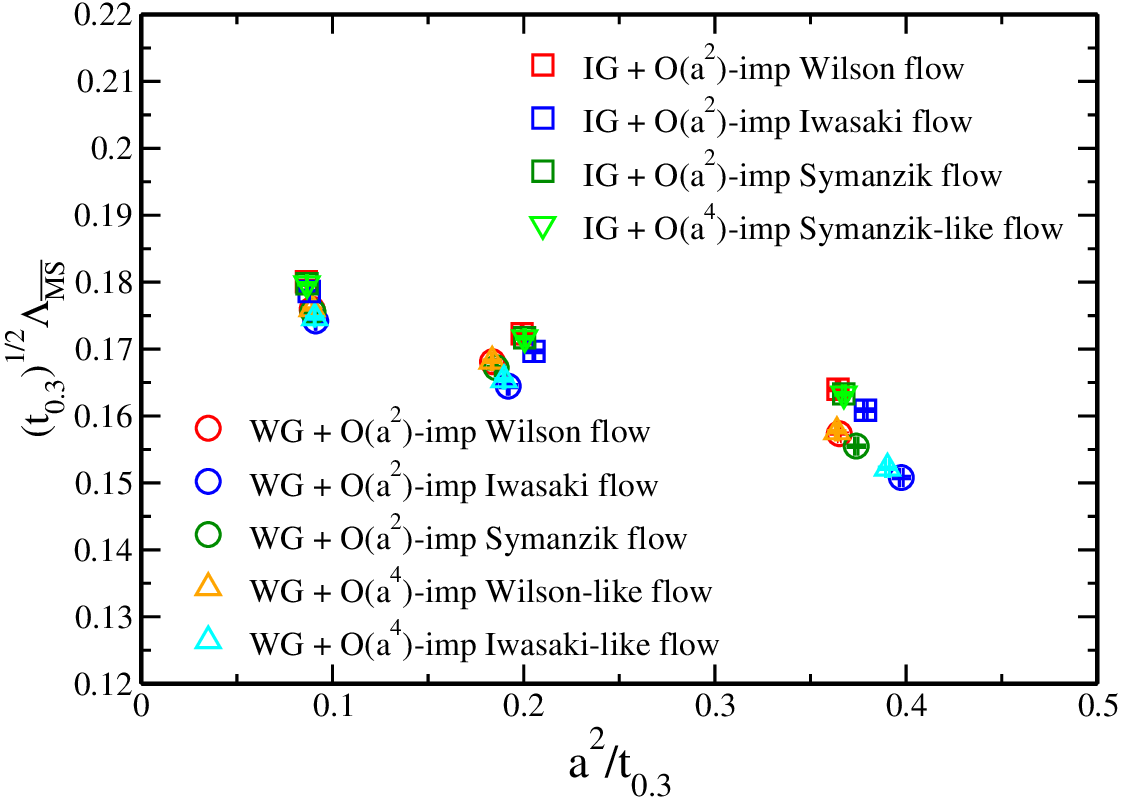}
\caption{The scaling behaviors of $\sqrt{t_X}\Lambda_{\overline{\textrm{MS}}}$
versus $a^2/t_X$ for choices of $X=0.15$ (left panels) and $X=0.3$ (right panels). 
The two upper panels show the results obtained by unimproved flows, while the two lower panels are for
the results obtained by $\mathcal{O}(a^2)$ and $\mathcal{O}(a^4)$-imp flows.
In the legends of these figures, WG (IG) stands for the Wilson (Iwasaki) gauge action
used in generating gauge configurations.
}
\label{fig:LambdaMSbar}
\end{figure*}
%

%
%
\begin{table}[ht]
\caption{
The one-loop and two-loop conversion variables $t_1^{\rm TI}$ and $t_2^{\rm TI}$
from the boosted lattice coupling to the $\overline{\textrm{MS}}$ coupling
for the standard Wilson ($c_{\rm rect}=0$) and the RG-improved Iwasaki ($c_{\rm rect}=-0.331$) 
gauge actions.~\label{tab:TIcoeff}
}
\begin{ruledtabular} 
\begin{tabular}{cllc}
$c_{\rm rect}$ & $t_1^{\rm TI}$ & $t_2^{\rm TI}$ & Ref.\cr
\hline
0 & $0.1348680$ & 0.0217565& \cite{Gockeler:2005rv}\cr
$-0.331$ & $0.1006$ & N/A & \cite{AliKhan:2001wr} \cr
\end{tabular}
\end{ruledtabular} 
\end{table}
%

%
%
\begin{table}[b]
\begin{ruledtabular} 
\caption{The expectation values of plaquette ($P$) and rectangle ($R$) values measured in this study.
\label{tab:PandR}
}
\begin{tabular}{lcccc}
 $\beta$ (Action) 
 &$P$ & $R$   \cr
\hline
5.96 (WG)
&  0.589 1583(32)& ---\cr
6.17 (WG)
&  0.610 8670(14)& ---\cr
6.42 (WG)
&  0.632 2170(06)  & --- \cr
\hline
2.60 (IG)
&  0.670 6232(15)& 0.452 8281(186)   \cr
2.80 (IG)
&  0.696 4317(61)& 0.490 0710(102)\cr
3.10 (IG)
&  0.727 6215(21)& 0.536 2049(037) \cr
\end{tabular}
\end{ruledtabular} 
\end{table}
%

%
%
\begin{table*}[ht]
\begin{ruledtabular} 
\caption{
Results of the continuum value of $\sqrt{t_X}\Lambda_{\overline{\rm MS}}$
for $X=0.15$ and 0.3 with various types of the tree-level improved flows.
The continuum extrapolation is performed by a least-square fit
to all three data points or two data points given at finer lattice spacings
using the linear form in terms of $a^2$.
\label{tab:ContinuumValues}
}
\begin{tabular}{lccccc}
type of calculation & type of extrapolation 
& $\sqrt{t_{0.15}} \Lambda^{\text{one-loop}}_{\overline{\textrm{MS}}}$ 
& $\sqrt{t_{0.15}} \Lambda^{\text{two-loop}}_{\overline{\textrm{MS}}}$ 
& $\sqrt{t_{0.3}} \Lambda^{\text{one-loop}}_{\overline{\textrm{MS}}}$ 
& $\sqrt{t_{0.3}} \Lambda^{\text{two-loop}}_{\overline{\textrm{MS}}}$ \cr
\hline
WG + $\mathcal{O}(a^2)$-imp Wilson flow & 3 points linear &  0.1176(1) & 0.1339(1) & 0.1813(2)& 0.2064(2) \cr
WG + $\mathcal{O}(a^2)$-imp Wilson flow & 2 points linear &  0.1189(1) & 0.1347(2) & 0.1834(3)& 0.2079(4)\cr
WG + $\mathcal{O}(a^4)$-imp Wilson-like flow & 3 points linear &  0.1175(1)& 0.1337(1)& 0.1812(2)& 0.2063(3)\cr
WG + $\mathcal{O}(a^4)$-imp Wilson-like flow & 2 points linear &  0.1188(1)& 0.1347(2)& 0.1834(3)& 0.2080(4)\cr
IG + $\mathcal{O}(a^2)$-imp Iwasaki flow  & 3 points linear & 0.1175(1) & N/A & 0.1832(2) & N/A\cr
IG + $\mathcal{O}(a^2)$-imp Iwasaki flow & 2 points linear  & 0.1191(1) & N/A & 0.1853(3) & N/A\cr
\end{tabular}
\end{ruledtabular} 
\end{table*}    
%

\subsection{Scaling behavior of the $\Lambda$ parameter}
\label{Sec:4-B}

In the following discussion, we use the one-loop formula in Eq.~(\ref{eq:matching_formula})
to evaluate the value of $g^2$ in the ${\overline{\rm MS}}$ scheme on both the Wilson and 
Iwasaki gauge configurations in order to treat them on the same footing, since
the value of $t_2^{\rm TI}$ is not known for the Iwasaki gauge action, as mentioned earlier.

In Fig.~\ref{fig:LambdaMSbar}, we plot the results for 
$\sqrt{t_X}\Lambda^{\text{one-loop}}_{\overline{\rm MS}}$
against $a^2/t_X$ from several combinations of two gauge actions and various flows
for choices of $X=0.15$ (left panels) and $X=0.3$ (right panels), 
as in Fig.~\ref{fig:sommer}. 
The two upper panels show the results obtained by unimproved flows, while the two lower panels are for
the results obtained by $\mathcal{O}(a^2)$ and $\mathcal{O}(a^4)$-imp flows.

For the unimproved flows (upper panels), the results for both cases of $X=0.15$ (left)
and 0.3 (right) very much depend on the choice of the gauge action and flow, similarly to the behavior of $\sqrt{t_X}/r_0$. 
However, the results given by all tree-level improved flows show a nearly perfect scaling behavior especially in the case of $X=0.15$ as a function of $a^2$ regardless of the types of 
gauge action and flow, unlike the behavior of $\sqrt{t_{X}}/r_0$. 
This indicates that the slight difference in the scaling behavior 
between the results obtained from WG and IG (which is found in Fig.~\ref{fig:sommer}) 
comes from the systematic uncertainties due to the scale setting by the Sommer scale.

For the case of $X=0.3$, the scaling behavior becomes less prominent. 
The behavior of $\sqrt{t_{0.3}}\Lambda_{\overline{\textrm{MS}}}$ 
indeed reveals a weak dependence of the choice of the gauge action, while 
the scaling behavior among various tree-level
improved flows on the same gauge configurations remains visible in the smaller region of $a^2$.
The violation of the scaling behavior of $\sqrt{t_{0.3}}\Lambda_{\overline{\textrm{MS}}}$ 
as a function of $a^2$ is mainly attributed to the fact that both Eqs.~(\ref{eq:matching_formula}) and (\ref{eq:MSbar_para})
are used beyond their applicability region.  Indeed, $t_{0.3}$ is located in the larger $t$ region, 
where the renormalized coupling $g^2$ becomes large, as shown 
in Figs.~\ref{fig:beta617}, \ref{fig:o4imp}, \ref{fig:beta280}, and \ref{fig:IG_o2imp}.

Even for the case of $\sqrt{t_{0.15}}\Lambda_{\overline{\textrm{MS}}}$, 
where the nearly perfect
scaling is achieved, there is still a slight linear dependence of $a^2$. 
However, if one reads off the 
slopes of the scaling behaviors from the lower left and right panels of 
Fig.~\ref{fig:LambdaMSbar}, the slope for $X=0.15$ is less steep than for $X=0.3$.
The origin of linear scaling in terms of $a^2$ is undoubtedly related to the 
remnant $\mathcal{O}(g^{2n} a^2)$ corrections, as we will discuss in detail later.

We now consider the continuum limit of the values of $\sqrt{t_{X}}
\Lambda_{\overline{\textrm{MS}}}$.
Among the various flow results, we focus on the tree-level $\mathcal{O}(a^2)$ improvement flows
for the following specific cases ($c_g=c_f$): the $\mathcal{O}(a^2)$-imp 
Wilson flow on the Wilson gauge configurations, and the $\mathcal{O}(a^2)$-imp 
Iwasaki flow on the Iwasaki gauge configurations.

For the continuum extrapolation, we simply adopt a linear form in terms of $a^2$:
%
%
\begin{equation}
\sqrt{t_X}\Lambda_{\overline{\rm MS}}(a)=\left(\sqrt{t_X}\Lambda_{\overline{\rm MS}}\right)_{\rm con}
+D_X \cdot \frac{a^2}{t_X}.
\label{eq:Fitform}
\end{equation}
Making a least-squares fit to all three data points using Eq.~(\ref{eq:Fitform}), we get
\begin{eqnarray}
\left(\sqrt{t_{0.15}}\Lambda^{\text{one-loop}}_{\overline{\rm MS}}\right)_{\text{con}}&=& 0.1176(1)\cr
\left(\sqrt{t_{0.3}}\Lambda^{\text{one-loop}}_{\overline{\rm MS}}\right)_{\text{con}}&=&   0.1813(2) \nonumber
\end{eqnarray}
from the WG + $\mathcal{O}(a^2)$-imp Wilson flow and
\begin{eqnarray}
\left(\sqrt{t_{0.15}}\Lambda^{\text{one-loop}}_{\overline{\rm MS}}\right)_{\text{con}}&=& 0.1175(1)\cr
\left(\sqrt{t_{0.3}}\Lambda^{\text{one-loop}}_{\overline{\rm MS}}\right)_{\text{con}}  &=& 0.1832(2)
\nonumber
\end{eqnarray}
from the IG + $\mathcal{O}(a^2)$-imp Iwasaki flow. 
Even if we exclude the data point at the coarsest lattice spacing from the fit, 
the results are not much different, as summarized in Table~\ref{tab:ContinuumValues}.
Clearly, for the smaller $X$ the resulting continuum value is stable against
the choice of the gauge action and the flow. 
In this sense, after the tree-level improvement is achieved, 
the reference scale $t_{0.15}$ is much better controlled in comparison to
the original one $t_{0.3}$. 

Recall that we do not take into account the possible
large systematic error stemming from the uncertainty 
of determining $\Lambda_{\overline{\rm MS}}$.
A precise determination of the continuum value of $\sqrt{t_{X}}\Lambda_{\overline{\textrm{MS}}}$ is 
the beyond scope of the present paper. The reason why we examine the scaling behavior of 
$\sqrt{t_{X}}\Lambda_{\overline{\textrm{MS}}}$ is that we would like to know
the scaling property purely obtained from the reference scale $t_{X}$
without unknown systematic uncertainties of the lattice spacing discretization errors,
arising from the introduction of other observables such as the Sommer scale $r_0$.

Next, we examine the uncertainty of determining $\Lambda_{\overline{\rm MS}}$. 
For the Wilson gauge action, the fully NLO formula 
of the conversion from the boosted coupling to
the $\overline{\rm MS}$ coupling is known with the values of $t_1^{\rm TI}$ and $t_2^{\rm TI}$
as given in Table~\ref{tab:TIcoeff}. 
Therefore, we would like to compare the results for $(\sqrt{t_{X}}\Lambda_{\overline{\textrm{MS}}})_{\rm con}$, 
which are determined with both the one-loop and two-loop conversions of Eq.~(\ref{eq:matching_formula}).
When the two-loop conversion is used, we get 
\begin{eqnarray}
\left(\sqrt{t_{0.15}}\Lambda^{\text{two-loop}}_{\overline{\rm MS}}\right)_{\text{con}}&=&0.1339(1)\cr
\left(\sqrt{t_{0.3}}\Lambda^{\text{two-loop}}_{\overline{\rm MS}}\right)_{\text{con}}&=&  0.2064(2) \nonumber
\end{eqnarray}
which indicates that the uncertainties stemming from the scheme conversion on the
coupling are estimated as about 12\% for the determination
of $(\sqrt{t_{X}}\Lambda_{\overline{\textrm{MS}}})_{\rm con}$.

\begin{figure}[ht]
\includegraphics[width=.40\textwidth,clip]{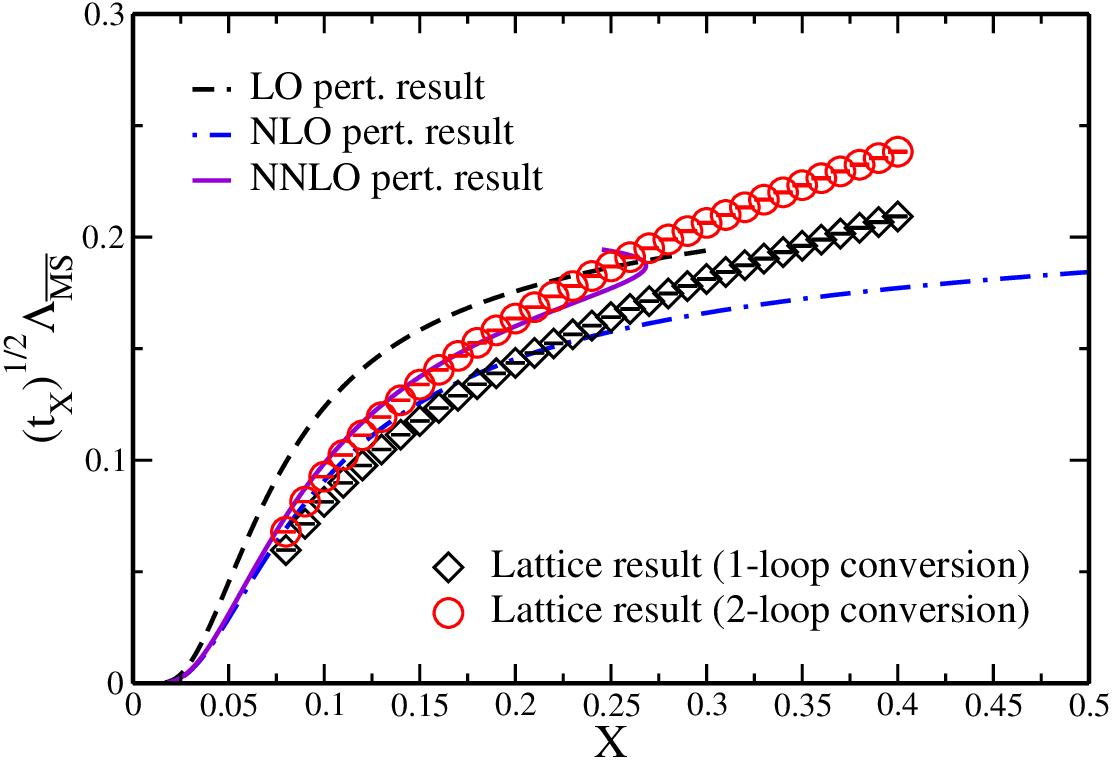}
\caption{Comparison of the lattice gradient flow and the perturbative gradient flow
 with respect to the values of $\sqrt{t_{X}}\Lambda_{\overline{\textrm{MS}}}$ for various choices of $X$.
For the lattice gradient flow, the continuum extrapolation is performed at fixed $X$.
The open circle (diamond) symbols represent the lattice results obtained with the 
$\Lambda_{\overline{\textrm{MS}}}$ parameter using the one-loop (two-loop) conversion 
from the lattice coupling to the $\overline{\textrm{MS}}$ coupling. 
The dashed, dot-dashed, and solid curves represent the LO, NLO, and NNLO results
from the perturbative gradient flow with the four-loop $\overline{\textrm{MS}}$
running coupling.
}
\label{fig:LQCD_vs_PQCD}
\end{figure}

In Fig.~\ref{fig:LQCD_vs_PQCD}, we plot the continuum extrapolated values
of $\sqrt{t_{X}}\Lambda_{\overline{\textrm{MS}}}$ 
for various choices of $X$ as a function of $X$ 
in the range of $0.1 \alt X \alt 0.4$. The results are obtained from data for the 
WG + $\mathcal{O}(a^2)$-imp Wilson flow. The open circle symbols represent
the results obtained when using the two-loop conversion, while the open diamond symbols
represent the results obtained when using the one-loop conversion. 
Clearly, the difference between two results for each $X$ is considerably larger 
than their own statistical errors; meanwhile, the difference becomes more pronounced for larger values of $X$. 

Figure~\ref{fig:LQCD_vs_PQCD} also includes three results from the continuum perturbation theory.
At a given scale $1/\sqrt{8t_X}$, we evaluate the four-loop $\overline{\textrm{MS}}$ running 
coupling~\cite{{vanRitbergen:1997va},{Czakon:2004bu}} and then calculate the value of $t_X\langle E(t_X)\rangle$ 
using the LO, NLO, and NNLO gradient flow formula of Eq.~(\ref{eq:flowed_energy}).
The dashed, dot-dashed and solid curves represent the LO, NLO, and NNLO results. 
Our numerical results for $(\sqrt{t_{X}}\Lambda^{\text{two-loop}}_{\overline{\textrm{MS}}})_{\rm con}$
from the lattice gradient flow are fairly consistent with the NNLO result from the perturbative gradient flow 
in the range of $0.1 \alt X \alt 0.2$. 

%
%
%

%
%
\begin{figure}[ht]
\includegraphics[width=.40\textwidth,clip]{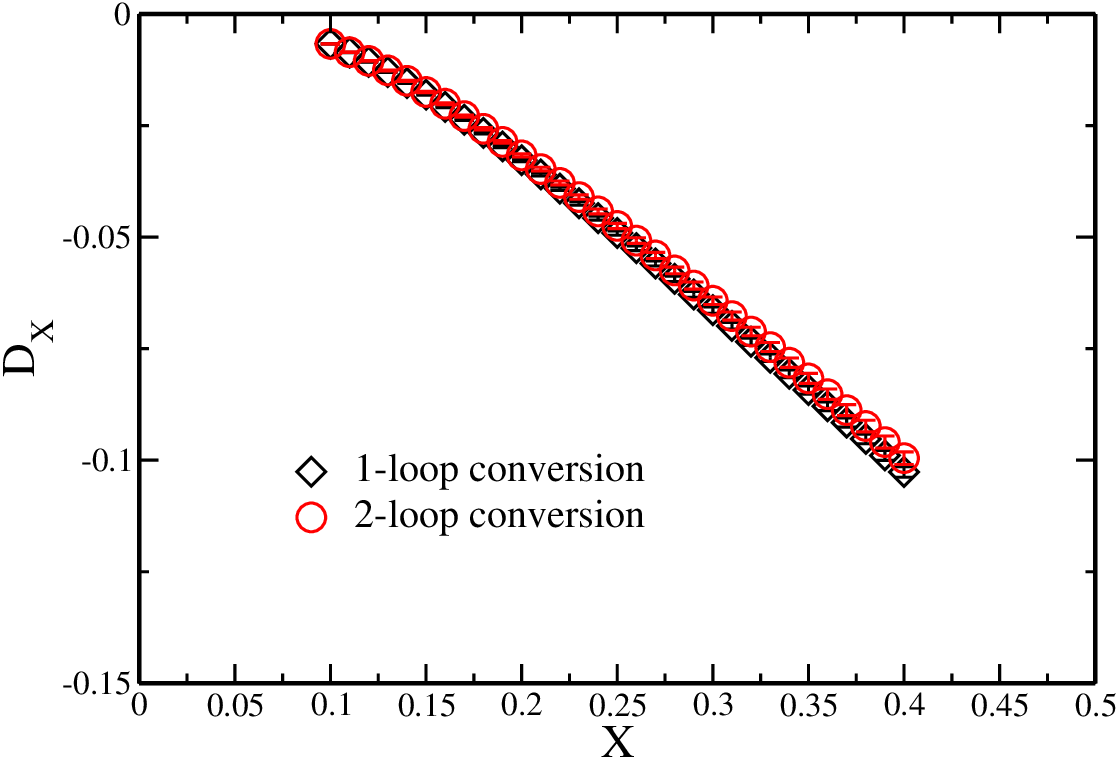}
\caption{The behavior of the coefficient $D_X$ as a function of $X$. 
$D_X$ is the coefficient of the linear term with respect to $a^2/t_X$ in 
the fit form of Eq.~(\ref{eq:Fitform}).
}
\label{fig:Slope_Xdep}
\end{figure}
%

%
%
\begin{figure}[ht]
\includegraphics[width=.40\textwidth,clip]{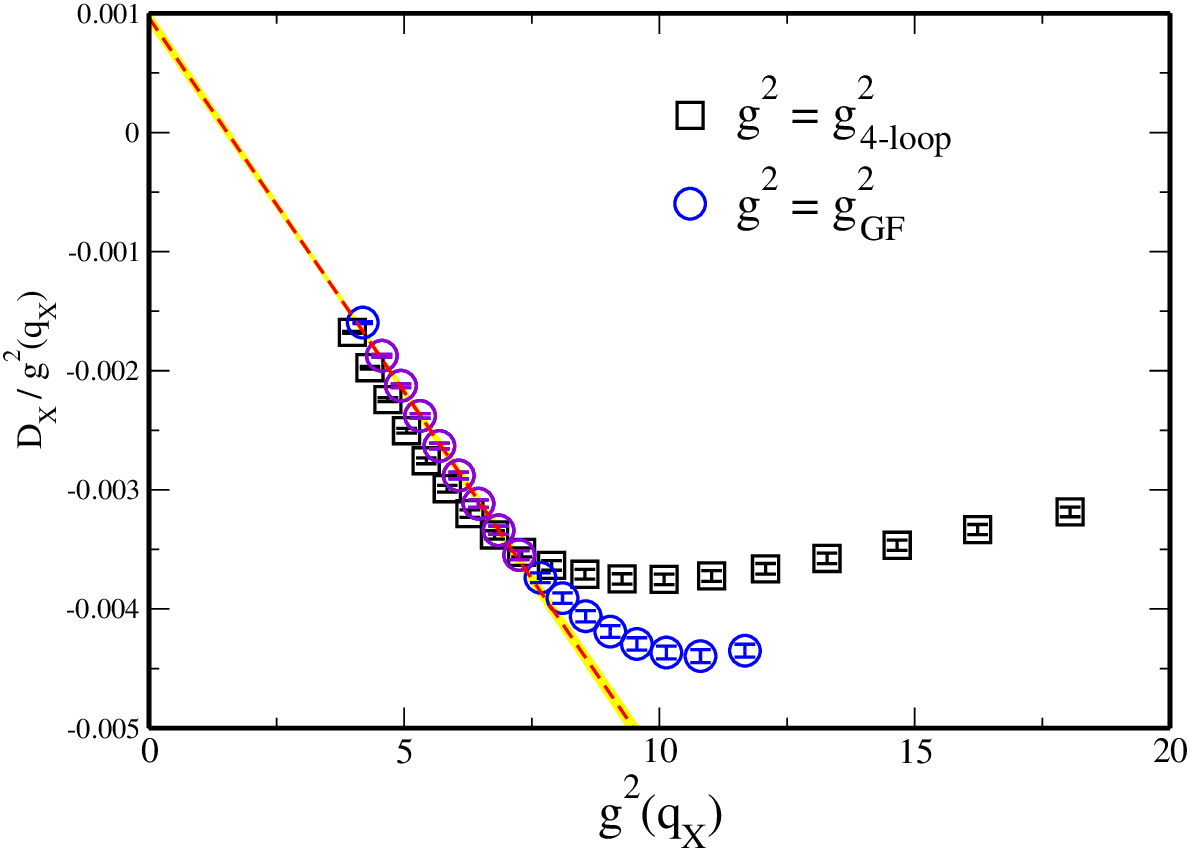}
\caption{The ratio of $D_X$ to $g^2$ as a function of $g^2$. The value of $g^2$
is evaluated by two methods---the four-loop running coupling (open squares)
and the GF coupling (open circles)---which are defined in the text. 
The dashed line represents the fit result to the data points given by the GF coupling 
with the linear form in terms of $g^2$.
}
\label{fig:g2a2_dep}
\end{figure}
%

\subsection{Remnant $\mathcal{O}(g^{2n} a^2)$ corrections}
\label{Sec:4-C}

Finally, we discuss the origin of linear scaling in terms of $a^2$ observed in Fig.~\ref{fig:LambdaMSbar}.
The strength of the remnant scaling violation that is approximately proportional to $a^2$ can 
be read off from the $X$ dependence of the coefficient $D_X$ defined in the fitting form 
of Eq.~(\ref{eq:Fitform}). In Fig.~\ref{fig:Slope_Xdep}, we show the $X$ dependence of 
the values of $D_X$ evaluated from both results when using the one-loop and two-loop conversions of 
Eq.~(\ref{eq:matching_formula}). Although there are large uncertainties 
in the determination of $(\sqrt{t_{X}}\Lambda_{\overline{\textrm{MS}}})_{\rm con}$
due to the choice of the conversion formula (\ref{eq:matching_formula}), 
the differences in the slope
coefficients $D_X$ obtained from two conversions are negligible
especially for the smaller $X$ region.

Therefore, in the following discussion we use the results for $D_X$ 
obtained from the two-loop conversion with the WG + $\mathcal{O}(a^2)$-imp Wilson flow.
As briefly mentioned earlier, a nonzero value of $D_X$ for a given $X$ (which is 
associated with the scaling violation) could be mainly caused by non-negligible 
$\mathcal{O}(g^{2n} a^2)$ corrections beyond the tree-level discretization effects, 
as shown in Fig~\ref{fig:Slope_Xdep}.
It indeed seems that the value of $D_X$ goes to zero as $X$ decreases. If this is true, 
$D_X$ might vanish because of the asymptotic freedom ($g^2 \rightarrow 0$) 
at high energies ($X\rightarrow 0$). Therefore, we assume
that $D_X$ is expressed by a power series in the $\overline{\textrm{MS}}$
running coupling $g$ at a scale of $q_X=1/\sqrt{8t_X}$ as follows:
%
%
\begin{equation}
D_X=\sum_{n} \frac{d_{2n}}{(4\pi)^n} \cdot g^{2n}(q_X)
\label{eq:gExpansion}
\end{equation}
where $d_{2n}$ are the perturbative expansion coefficients. 
If the summation in the above equation starts at $n=1$ rather than $n=0$,
the origin of linear scaling is mainly associated with the remnant 
$\mathcal{O}(g^{2n} a^2)$ corrections, which arise beyond the tree level.

To verify this assumption, let us plot the ratio of $D_X$ to $g^2$ as a function of $g^2$,
as shown in Fig.~\ref{fig:g2a2_dep}. As for the value of $g^2$, we consider two types of
estimation. 1) One type is to use the perturbative four-loop expression for the $\overline{\textrm{MS}}$
running coupling $g^2(q_X)$ with our numerical results for
$(\sqrt{t_{X}}\Lambda^{\text{two-loop}}_{\overline{\textrm{MS}}})_{\rm con}$ at a given $X$.
2) Another type is to use the gradient flow (GF) coupling $g^2_{\textrm{GF}}$~\cite{Fodor:2012td},
which can be determined from the gradient flow formula of Eq.~(\ref{eq:flowed_energy})
with a given value of $t^2\langle E(t)\rangle$. In the latter, we adopt the NNLO formula, which yields 
a cubic equation with respect to $g^2(q_X)$ at a fixed $X$.
We thus evaluate the ratio of $D_X$ to $g^2$ using the above two methods,
as shown in Fig.~\ref{fig:g2a2_dep}.

The open square symbols represent the ratio given by the four-loop running coupling, while
the open circle symbols are evaluated using the GF coupling. When  $g^2 \agt 4\pi$, 
the cubic equation with respect to $g^2$ admits no feasible solution. 
For each case, we thus show 17 data points, which are in the range of $0.10 \le X \le 0.28$.
In the weaker coupling regime ($g^2\alt 7$), two results from different evaluations 
of the ratio $D_X/g^2$ overlap each other. Their eight or nine data points ($0.1 \le X \alt 0.18$) 
start to show the expected weak coupling scaling behavior, which is almost linear in $g^2$. 
To ensure this point, we have carried out the linear fit on the data set given by
the GF coupling, which exhibits a milder $g^2$ dependence,  using the following expression:
%
%
\begin{equation}
D_X/g^2(q_X)=\frac{d_2}{4\pi} + \frac{d_4}{(4\pi)^2}\cdot g^2(q_X)
\label{eq:Fitform2}
\end{equation}
where $d_2$ and $d_4$ correspond to the coefficients for 
the first and second orders of $g^2$ in Eq.~(\ref{eq:gExpansion}).

The stability of the fit results has been tested against the number
of fitted data points. The best fit is drawn to fit eight data points, which are
indicated by violet open circles in Fig.~\ref{fig:g2a2_dep}, with a reasonable 
value of $\chi^2/{\rm d.o.f.}\approx 1.0$. We then obtain the results
{\begin{eqnarray}
 d_2&=& +1.2(1)\times 10^{-2}\cr
 d_4&=& -9.9(2) \times 10^{-2} \nonumber
\end{eqnarray}
}%
which indicate that the remnant $\mathcal{O}(g^{2n}a^2)$ corrections are 
reasonably small in the weak-coupling regime.
The fit result with one standard deviation is indicated by a red dashed line with a yellow shaded band
in Fig.~\ref{fig:g2a2_dep}. 
From the above observation regarding $D_X$, we conclude that the origin of linear scaling in terms of
$a^2$ found in Fig.~\ref{fig:LambdaMSbar} is related to the remnant $\mathcal{O}(g^{2n} a^2)$ 
corrections, which are beyond the tree level.
Although it is thus evident that the tree-level improvement program studied in this paper
is not enough to eliminate all $\mathcal{O}(a^2)$ effects,
the remnant $\mathcal{O}(g^{2n}a^2)$ corrections can be well under control even
when using the simple method for the tree-level $\mathcal{O}(a^2)$ improvement.


\begin{figure*}[ht]
\includegraphics[width=.32\textwidth,clip]{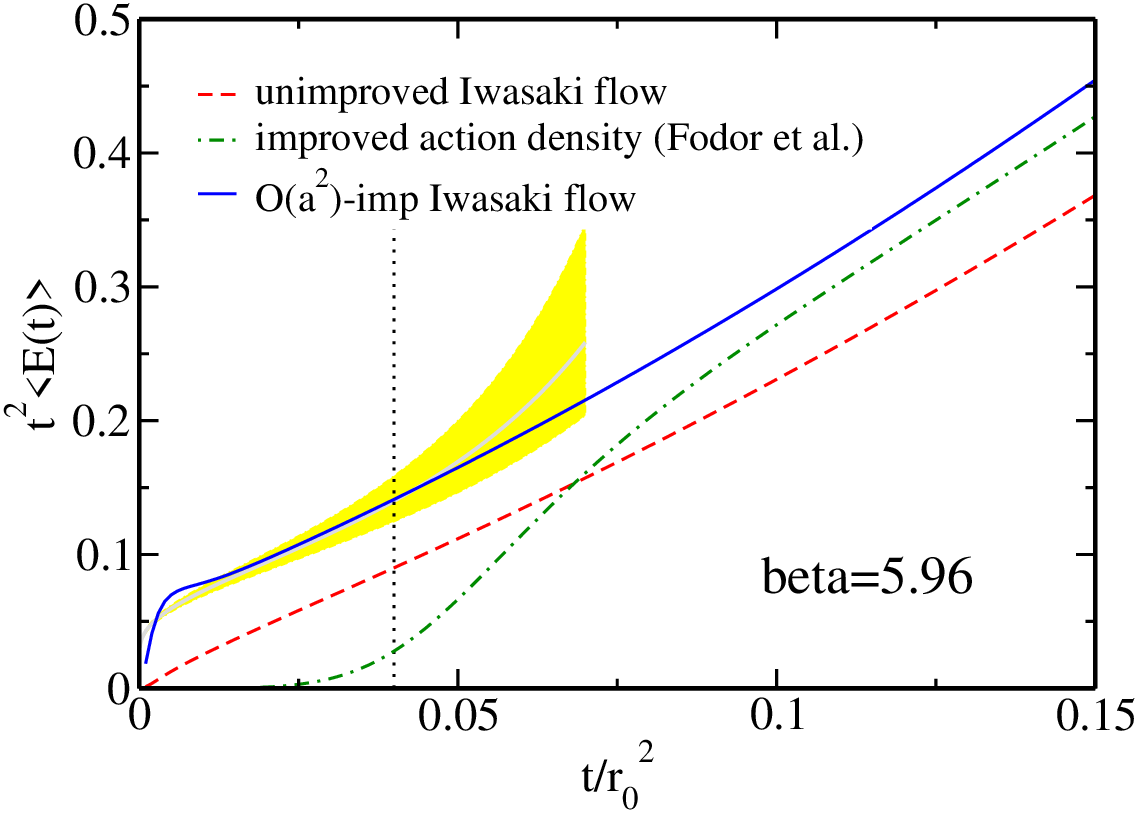}
\includegraphics[width=.32\textwidth,clip]{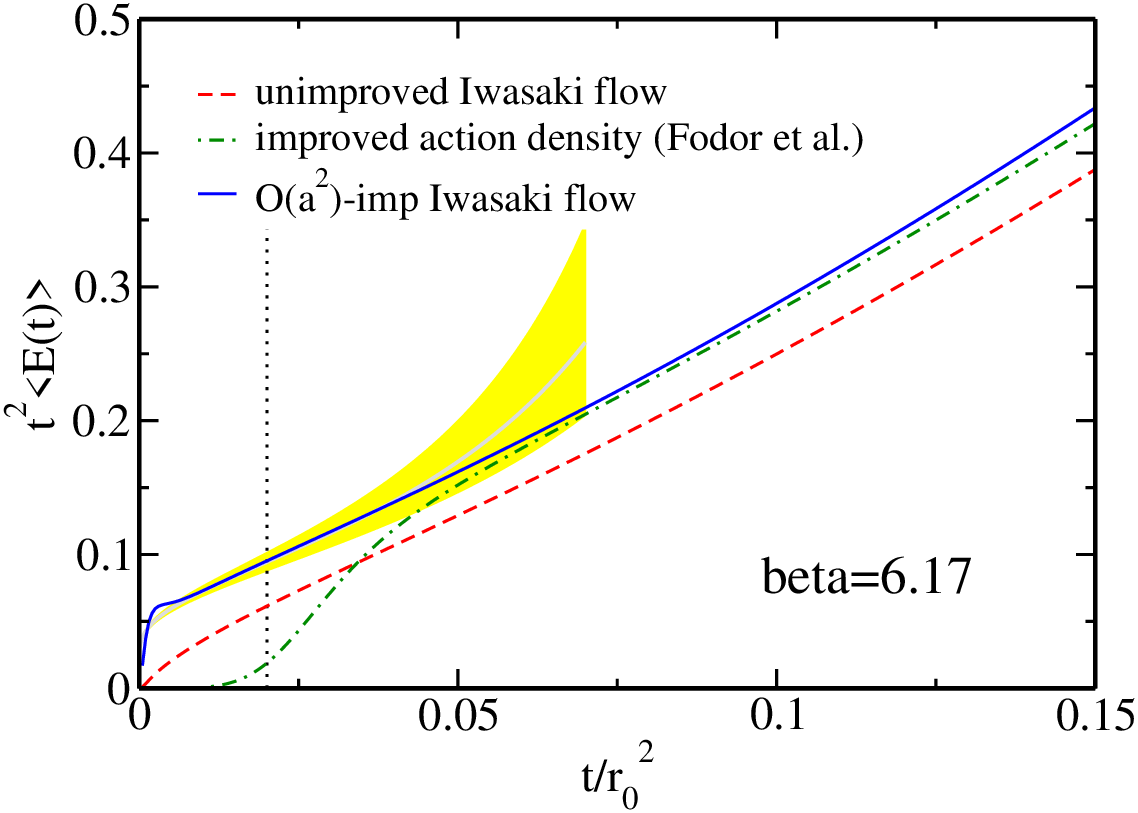}
\includegraphics[width=.32\textwidth,clip]{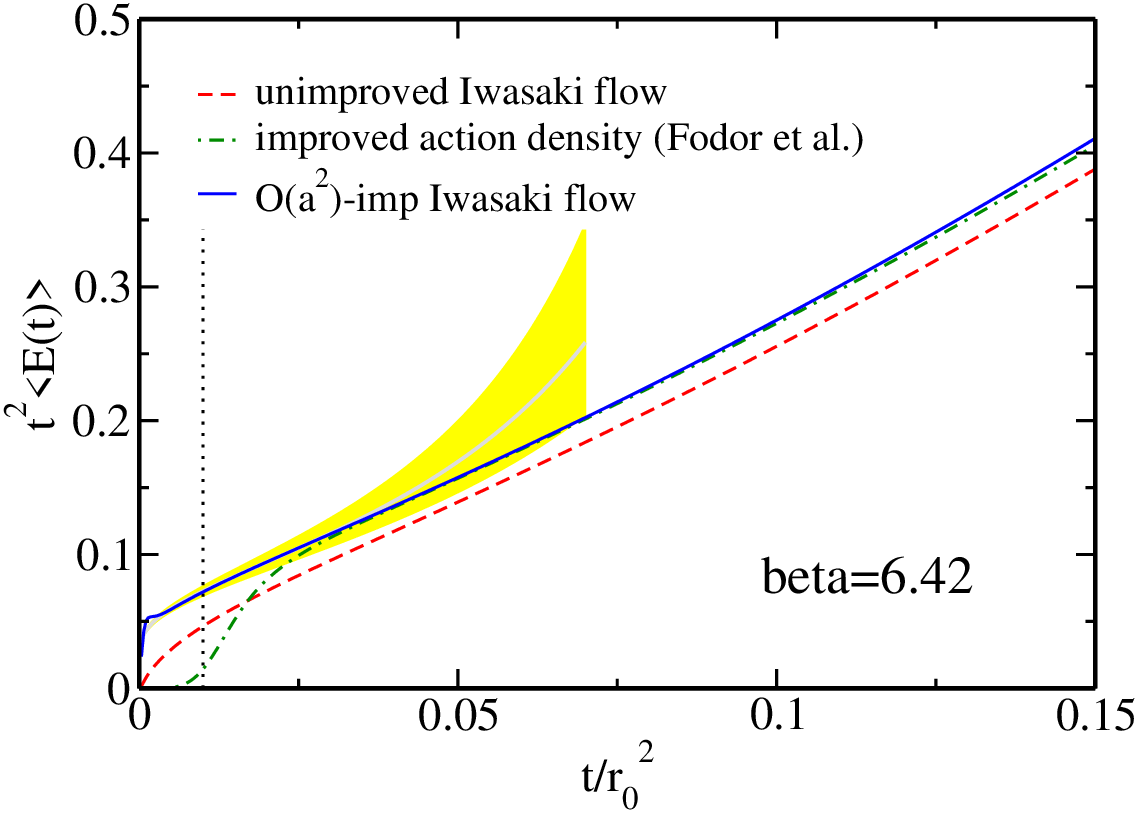}
 \caption{Comparisons with results obtained from the original proposal of the tree-level 
 improvement on the action density as defined in Eq.(\ref{eq:fodor_imp})~\cite{Fodor:2014cpa}.
 The results for $t^2\langle E(t)\rangle$ are calculated with the Iwasaki flow on the Wilson 
 gauge configurations. The three panels show the results calculated at $\beta=5.96$ (left), 
 $\beta=6.17$ (center), and $\beta=6.42$ (right).
 The unimproved results for the Iwasaki flow with the clover-type action density
 are represented by the red dashed curve in each panel, while their improved results given by
 Eq.(\ref{eq:fodor_imp}) are represented by the green dot-dashed curve 
 in each panel. The blue solid curve in each panel represents 
 our results obtained from the $\mathcal{O}(a^2)$-imp Iwasaki flow 
 (using the same graphical conventions as in Fig.~\ref{fig:beta617}.)}
 \label{fig:iwa_fodor}
 \end{figure*}

\section{Summary} 
\label{Sec:5}
We have studied several types of tree-level improvement on the Yang-Mills gradient flow
in order to reduce the lattice discretization errors on the expectation value 
of the action density $\langle E (t) \rangle$, in line with Ref.~\cite{Fodor:2014cpa}. 
For this purpose, the rectangle term was included in both the flow and gauge actions 
in the minimal way.
We proposed a simple idea of achieving tree-level $\mathcal{O}(a^2)$ and $\mathcal{O}(a^4)$ 
improvements on $\langle E (t) \rangle$, using the linear combination of two types of $\langle E (t) \rangle$ 
given by the plaquette- and clover-type definitions. To test our proposal, 
numerical simulations have also been performed with both the Wilson and Iwasaki 
gauge configurations generated at various lattice spacings.

Our numerical results have showed that tree-level lattice discretization errors 
on the quantity of $t^2\langle E (t) \rangle$ 
are certainly controlled in the small-$t$ regime for up to $t\agt a^2$
by both tree-level $\mathcal{O}(a^2)$- and $\mathcal{O}(a^4)$-improved flows. 
On the other hand, the values of $t^2\langle E (t) \rangle$ in the large-$t$ regime 
are different among the results given by different flow types, leading to the same 
improved flow up to either $\mathcal{O}(a^2)$ or $\mathcal{O}(a^4)$ at tree level.

In order to demonstrate the feasibility of our tree-level  improvement proposal, 
we first plotted the behavior of $\sqrt{t_X}/r_0$ as a function of $a^2/t_X$
in a similar manner as in Ref.~\cite{Luscher:2010iy}. 
However, the ratio $\sqrt{t_X}/r_0$ also contains the discretization errors in the 
determination of $r_0$ in addition to those of the lattice gradient flow.
We then studied the scaling behavior of the dimensionless combination of two scale parameters, 
$\sqrt{t_X}\Lambda_{\overline{\textrm{MS}}}$, which is free in the weaker 
coupling regime from unknown systematic uncertainties regarding the lattice spacing discretization errors 
arising from the introduction of other observables on the lattice, such as the Sommer scale.

For the smaller $X$, {\it e.g.}, $X=0.15$, once any tree-level improvement is achieved,
$\sqrt{t_{0.15}}\Lambda_{\overline{\textrm{MS}}}$ shows a nearly perfect scaling
behavior as a function of $a^2$ regardless the types of gauge action and flow. 
However, there is still a slight linear dependence of $a^2$ appearing in the cases of both 
$\mathcal{O}(a^2)$- and $\mathcal{O}(a^4)$-improved flows.
On the other hand, for the larger $X$, {\it e.g.}, the original choice of $X=0.3$, 
the behavior of $\sqrt{t_{0.3}}\Lambda_{\overline{\textrm{MS}}}$ reveals a weak dependence
of the choice of the gauge action, while the scaling behavior among various tree-level
improved flows on the same gauge configurations remains visible especially in the 
smaller region of $a^2$. 

All of the aforementioned features regarding the slight scaling 
violation and the gauge action dependence suggest that 
there still remain remnant 
$\mathcal{O}(g^{2n}a^2)$ corrections, 
which are beyond the tree level.
Indeed, the origin of linear scaling in terms of $a^2$ found in 
$\sqrt{t_{0.15}}\Lambda_{\overline{\textrm{MS}}}$ 
is related to the remnant $\mathcal{O}(g^{2n}a^2)$ corrections, which
can be read off from the $g^2$ dependence of the slope coefficient 
associated with the linear scaling. 

Although it is evident that the tree-level improvement program studied in this paper
is not enough to eliminate all $\mathcal{O}(a^2)$ effects, the remnant 
$\mathcal{O}(g^{2n}a^2)$ corrections can be well under control even
with the simple method for the tree-level $\mathcal{O}(a^2)$ improvement. 
Once the tree-level $\mathcal{O}(a^2)$ and $\mathcal{O}(a^4)$ improvements 
are achieved, the resulting energy density $\langle E(t)\rangle$ 
becomes very close to the continuum one in the small-$t$ regime for up to $t\agt a^2$. 
This offers an alternative reference scale $t_{X}$ with the smaller value of $X$, such as 
$X=0.15$. Indeed the continuum-extrapolated value of $t_{0.15}$ is 
in excellent agreement with the perturbative gradient flow result. 
On the other hand, it is observed that the original reference scale $t_{0.3}$ suffers from 
rather large $\mathcal{O}(g^{2n}a^2)$ errors when the lattice spacing is coarse, as large
as $a\approx 0.1$ fm.

\appendix
\section{Comparison with the original proposal}
\label{App:A}

Fodor {\it et al.} proposed a tree-level improvement of the action density 
$\langle E(t)\rangle$~\cite{Fodor:2014cpa} as defined in Eq.~(\ref{eq:fodor_imp}). 
In this appendix, we compare our results with results obtained from the original proposal.  
In Fig.~\ref{fig:iwa_fodor}, we show the results from the Iwasaki flow 
on the Wilson gauge configurations at three lattice spacings.
This particular combination of the gauge action and the flow provides 
large differences between the original proposal and ours. 

The three panels of Fig.~\ref{fig:iwa_fodor} show the results calculated at $\beta=5.96$ (left), 
$\beta=6.17$ (center), and $\beta=6.42$ (right).
The unimproved results for the Iwasaki flow with the clover-type action density
are represented by the red dashed curve in each panel, while their improved results given by
Eq.~(\ref{eq:fodor_imp}) are represented by the green dot-dashed curve 
in each panel. The blue solid curve in each panel represents our results
obtained from the $\mathcal{O}(a^2)$-imp Iwasaki flow.

Figure~\ref{fig:iwa_fodor} shows that the improved action density defined in 
Eq.~(\ref{eq:fodor_imp}) does not efficiently improve the behavior of $t^2\langle E(t)\rangle$
as a function of $t/r_0^2$ at the coarser lattice spacing, while our results
from the $\mathcal{O}(a^2)$-imp Iwasaki flow are much closer to the continuum perturbative calculation
even in the relatively small-$t$ regime up to $t/r_0^2\approx 0.02$, which is beyond 
the boundary of asymptotic power-series expansions in terms of $a^2/t$ at $\beta=5.96$.

The difference between the original proposal and ours becomes diminished as the lattice spacing decreases. 
Therefore, in results given by the original proposal the large deviation from the continuum perturbative 
calculation---which is found at the coarser lattice spacing---is certainly caused by the lattice discretization errors. 
Indeed, a simple division of the measured action density by the tree-level contribution $C(a^2/t)$
could not eliminate the tree-level discretization corrections properly unless $\sum _{n=1} ^{\infty} C_{2n} \cdot a^{2n} /t^n \ll 1$. 
This is the case when the large rectangle coefficient $c_{{\rm rect},f}=c_f$ is chosen for
the flow action. As shown in Secs.~\ref{Sec:3} and \ref{Sec:4}, our proposal does not have such 
a restriction on the value of $c_f$, and can equally eliminate the tree-level discretization corrections 
for nonzero values of $c_f$.

\begin{acknowledgments}
 We would like to thank the members of the FlowQCD Collaboration 
 (T. Hatsuda, T. Iritani, E. Itou, M. Kitazawa and H. Suzuki) for helpful suggestions and fruitful discussions.
 This work is in part based on the Bridge++ code (http://bridge.kek.jp/Lattice-code/) 
 and numerical calculations were partially carried out on supercomputer resources: 
 SR16000 and XC40 at YTIP, Kyoto University, SR16000 under the
Large-scale Simulation Program (No.15/16-02) at KEK, and LX406Re-2 under the
HPCI Systems Research Projects (Project ID: hp160020) at Cyberscience Center,
Tohoku University.

\end{acknowledgments}



\begin{thebibliography}{99}
\bibitem{difference}
Please note that this linear combination is different from the tree-level $\mathcal{O}(a^{2})$
``operator," $\frac{4}{3}\langle E_{\rm plaq}(t)\rangle - \frac{1}{3}\langle E_{\rm clover}(t)\rangle$,
proposed by Ramos and Sint~\cite{Ramos:2015baa}.
\bibitem{rectangle}
Note that although the positive rectangle flow yields positive values of 
$\langle E_{\rm plaq} (t) \rangle$ and $\langle E_{\rm clover} (t) \rangle$,
the $\mathcal{O}(a^4)$-improved combination of Eq.(\ref{eq:combination}) makes
$\langle E (t)\rangle$ negative.
\end{thebibliography}
\end{document}